\documentclass[reprint,onecolumn,superscriptaddress,showpacs,nofootinbib,notitlepage]{revtex4-1}

\usepackage{graphicx}
\usepackage{latexsym,amsmath,amssymb,lmodern,float,url}
\usepackage{natbib}
\usepackage{color}
\usepackage{microtype}
\usepackage{slashed}
\usepackage{multirow}

\newcommand{\be}{\begin{equation}}
\newcommand{\ee}{\end{equation}}
\newcommand{\bea}{\begin{eqnarray}}
\newcommand{\eea}{\end{eqnarray}}
\newcommand{\eq}[1]{Eq.~(\ref{eq:#1})}
\newcommand{\figu}[1]{Fig.~\ref{fig:#1}}
\newcommand{\tab}[1]{Table~\ref{tab:#1}}

\newcommand{\nn}{\nonumber}
\newcommand{\ba}{\begin{array}}
\newcommand{\ea}{\end{array}}

\newcommand{\re}{\mathop{\rm{Re}}}
\newcommand{\im}{\mathop{\rm{Im}}}

\usepackage[colorlinks=true,backref=false, linktocpage=true,
citecolor=blue,urlcolor=blue,linkcolor=blue,pdfpagemode=UseOutlines]{hyperref}

\hypersetup{%
  bookmarksnumbered=true,
  pdftitle = {},
  pdfsubject = {},
  pdfauthor = {},
  pdfkeywords = {}
}

\def\Re {\mathop{\hbox{Re}}}
\def\Im {\mathop{\hbox{Im}}}

\begin{document}
\title{ Deep Learning Beyond Lefschetz Thimbles}

\author{Andrei Alexandru }
\email{aalexan@gwu.edu}
\affiliation{Department of Physics, The George Washington University, Washington, D.C. 20052, USA}
\affiliation{Department of Physics, University of Maryland, College Park, MD 20742, USA}
\affiliation{Albert Einstein Center for Fundamental Physics, Institute for Theoretical Physics, University of Bern, Sidlerstrasse 5, CH-3012 Bern, Switzerland}
\author{Paulo F. Bedaque}
\email{bedaque@umd.edu}
\affiliation{Department of Physics, University of Maryland, College Park, MD 20742, USA}
\author{Henry Lamm}
\email{hlamm@umd.edu}
\affiliation{Department of Physics, University of Maryland, College Park, MD 20742, USA}
\author{Scott Lawrence}
\email{srl@umd.edu}
\affiliation{Department of Physics, University of Maryland, College Park, MD 20742, USA}
\date{\today}

\begin{abstract}
The generalized thimble method to treat field theories with sign problems requires repeatedly solving the computationally-expensive holomorphic flow equations. We present a machine learning technique to bypass this problem. The central idea is to obtain a few field configurations via the flow equations to train a feed-forward neural network. The trained network defines a new manifold of integration which reduces the sign problem and can be rapidly sampled. We present results for the $1+1$ dimensional Thirring model with Wilson fermions on sizable lattices.
In addition to the gain in speed, the parameterization of the integration manifold we use avoids the ``trapping" of Monte Carlo chains which plagues large-flow calculations, a considerable shortcoming of the previous attempts.
\end{abstract}

\maketitle\section{Introduction}

Monte Carlo methods are widely used in the study of field theoretical and many-body systems.
They can be understood as a way of computing a large dimensional integral encoding the physics of the system through the importance sampling of the integrand. In field theories, this integral is a discretized version of the Feynman path integral. The Monte Carlo method is essentially the only general purpose method capable of dealing with strongly interacting field theories. Unfortunately, many systems of great interest are defined by path integrals where the integrand oscillates wildly, making a direct stochastic estimation impossible in practice. This situation is referred to as the ``sign problem". The class of systems with a severe sign problem  includes most finite-density systems, among them QCD, condensed matter models (e.g. the Hubbard model away from half-filling), and all real-time calculations. 
In the context of QCD,
many ideas aiming at solving the sign problem have been developed through the years, among them the complex Langevin method \cite{Aarts:2008rr}, the density of states method  \cite{Langfeld:2016mct}, canonical methods \cite{Alexandru:2005ix,deForcrand:2006ec}, reweighting methods \cite{Fodor:2001au}, series expansion on the chemical potential \cite{Allton:2002zi}, fermion bags \cite{Chandrasekharan:2013rpa} and analytic continuation from imaginary chemical potentials \cite{deForcrand:2006pv}.

In the last few years the thimble approach \cite{Cristoforetti:2012su,Cristoforetti:2013qaa} has received a lot of attention. The main idea of this method is to deform the region of integration of the path integral from the original real fields to some other manifold, $\mathcal{M}$, embedded in the space of complexified fields. This deformation, if care is exercised, does not change the value of the integral, thanks to a multidimensional generalization of the Cauchy theorem of complex analysis. If $\mathcal{M}$ is properly chosen, the sign problem can be solved or, at least, substantially reduced.

In the first attempts, $\mathcal{M}$ was chosen to be the set of Lefschetz thimbles, which are the multidimensional generalization of the ``steepest descent" or ``constant phase" paths of complex analysis \cite{Cristoforetti:2014gsa,DiRenzo:2015foa,Mukherjee:2013aga,Fujii:2015vha,Fukushima:2015qza,Alexandru:2015xva}. The phase of the (Euclidean) action, and consequently of the Boltzmann factor $e^{-S}$,  is constant over one thimble\footnote{An additional contribution to the integral from the curvature of the thimble can be nonzero, but this ``residual phase'' is typically found to be small.}, so the sign problem is ameliorated. Two complications immediately appear. The first is that the thimbles cannot be found analytically in non-trivial models so an algorithm to compute them is necessary to perform the integral. This is a non-trivial task, but a few proposals have been put forward \cite{Cristoforetti:2014gsa,Fukushima:2015qza,Alexandru:2015xva}. The second complication is that it is very hard to determine which combination of thimbles equals the original integral over real fields. It has been conjectured that one thimble dominates the path integral in the thermodynamic and/or continuum limits \cite{Cristoforetti:2012su} while, at the same time, the importance of multiple thimbles has been demonstrated  in finite volume systems \cite{Alexandru:2015sua,Tanizaki:2015rda}.

A new method, inspired by the thimble approach, was suggested in \cite{Alexandru:2015sua}. The deformation of the manifold of integration $\mathcal{M}_T$ is chosen to be  the result of evolving every point $\zeta$ of the original integration region (the set of real fields, identified with $\mathbb{R}^N \subset \mathbb C^N$) through the holomorphic flow equations by a ``time" $T$:
\bea \label{eq:flow}
\frac{d\phi_i}{dt} = \overline{\frac{\partial S}{\partial \phi_i}},\, \phi_i(0) = \zeta_i, 
\eea 
where $S$ is the (Euclidean) action of the model and the bar denotes complex conjugation. The point $\tilde\phi$ in $\mathcal{M}_T$ corresponding to $\zeta$ is given by $\tilde\phi(\zeta) = \phi(T)$.
As the flow time $T$ increases, the set of all points thus obtained ($\mathcal{M}_T$) approaches the right combination of thimbles which equals the original integral over $\mathbb{R}^N$. At intermediate values of flow time $T$,  $\mathcal{M}_T$ differs from the thimbles and the sign problem is not completely solved but ameliorated.  For this reason, the method introduced in \cite{Alexandru:2015sua} is sometimes called the ``generalized thimble" approach \cite{Nishimura:2017vav}. 
This approach has the advantage that neither {\it a priori} information about the correct combination of thimbles equivalent to the integration over real fields, nor their location and shape,  is required. This method was demonstrated in two dimensional models both in Euclidean time \cite{Alexandru:2016ejd} and in real time (Minkowski space) \cite{Alexandru:2016ejd}. However, as the method was applied to larger systems two complications became apparent. They are a consequence of the fact that, in some models and parameter values, the sign problem is alleviated only after considerable flow. The first problem  is that large flow times are computationally expensive. The second is that the probability distribution over $\mathbb R^N$ --- induced by the probability distribution over $\mathcal M_T$ and the parameterization $\phi(\zeta)$ --- typically becomes strongly multimodal, with wide barriers of low  probability separating isolated regions of high probability. Such a distribution is difficult to properly sample through local updates. Although methods to deal with the multimodality have been proposed \cite{Fukuma:2017fjq,Alexandru:2017oyw}, they significantly increase the computational cost of the calculation.

This paper introduces a substantial add-on to the generalized thimble method which addresses the two shortcomings described above. The main idea is to use a parameterized form of the manifold $\mathcal{M}_T$ obtained by interpolating between some (complex) fields obtained from evolving real configurations by   \eq{flow}. This multidimensional  interpolation is a complex non-linear regression problem, a very non-trivial task. We approach it using machine learning techniques. More precisely, we will use a feed-forward network which inputs a real configuration $\phi_R$ and outputs the corresponding imaginary part on the integration manifold:
\be
\phi_R \rightarrow \tilde\phi=\phi_R + i \tilde{f}(\phi_R),
\ee where the function $\tilde f$ is implemented using the feed-forward network. 
The network is ``trained" in such a way that the set of complex fields $\tilde\phi$ obtained by running all real $\phi$, the ``learnifold" $\mathcal{L}_T$, approximates the flowed manifold $\mathcal{M}_T$.  The advantage of using the network to generate the configurations is that it bypasses the computationally expensive, repeated solution of \eq{flow} (and the even more expensive computation of a Jacobian, see below). In addition, the parameterization $\tilde\phi=\phi_R + i \tilde{f}(\phi_R)$ has better properties compared to the one used previously regarding the multimodality problem, as explained below.

We will review the generalized thimble method in Sec.~\ref{sec:gtm} and the use of feed-forward network methods in Sec.~\ref{sec:ffn}. The specifics of the learnifold will be covered in Sec.~\ref{sec:learnifold}. In Sec.~\ref{sec:model} we discuss the $1+1$ dimensional Thirring model, which we will the use to test and demonstrate our method. The details of the simulations are presented in Sec.~\ref{sec:algorithm} and results are presented in Sec.~\ref{sec:results}. Conclusions are summarized in Sec.~\ref{sec:con}.

\section{Generalized thimble method}\label{sec:gtm}

The expectation value of an observable in field theory can be cast in the form of a path integral
\be
\langle\mathcal{O}\rangle = \frac
{
\int_{\mathbb{R}^N} D\phi\; e^{-S[\phi] }\mathcal{O}[\phi]
}
{
\int_{\mathbb{R}^N} D\phi\; e^{-S[\phi] }
},
\ee where $S[\phi]$ is the (Euclidean) action. The stochastic evaluation of this ratio is accomplished by approximating it by
\be
\langle\mathcal{O}\rangle \approx \frac{1}{N_s} \sum_{\phi^{(s)}} \mathcal{O}[\phi^{(s)}],
\ee where the configurations $\phi^{(s)}$ are sampled randomly from the distribution $p[\phi] \sim e^{-S[\phi]}$.  The exact value is approached as the number of configurations $N_s$ grows. 

This method works if $S[\phi]$ is real; otherwise the sampling can be done with respect to $\sim e^{-\re S[\phi]}$ and the phase of the integrand included when computing observables by reweighting
\begin{align}\label{eq:realint}
\langle{\mathcal{O}} \rangle &= \frac
{
\int_{\mathbb{R}^N} D\phi\; e^{-S[\phi] } \mathcal{O}[\phi]
}
{
\int_{\mathbb{R}^N} D\phi\; e^{-S[\phi] }
}
\nn\\&=
 \frac
{
\int_{\mathbb{R}^N} D\phi\; e^{-S_R[\phi] }  \mathcal{O}[\phi] e^{-i S_I[\phi] } 
}
{
\int_{\mathbb{R}^N} D\phi\; e^{-S_R[\phi] }
}
\frac
{
\int_{\mathbb{R}^N} D\phi\; e^{-S_R[\phi] }  }
{
\int_{\mathbb{R}^N} D\phi\; e^{-S_R[\phi] } e^{-i S_I[\phi] } 
}
\nn\\&=
\frac{\langle \mathcal{O} e^{-i S_I }   \rangle_{S_R}}{\langle e^{-i S_I } \rangle_{S_R}}.
\end{align}
with $S_R=\re S, S_I=\im S$. This procedure is practical only if the average phase $\langle e^{-i S_I }\rangle_{S_R}$ is large. Otherwise the calculated expectation value will result from a ratio of small numbers, each resulting from detailed cancellations among configurations. In most theories the average sign is expected to vanish exponentially as the volume increases and/or the temperature decreases. This is the sign problem.

 The idea of the generalized thimble method is to deform the domain of integration from $\mathbb{R}^N$ to $\mathcal{M}_T$, a submanifold of $\mathbb{C}^N$ of (real) dimension $N$:

\begin{align}\label{eq:comint}
\langle\mathcal{O}\rangle&= \frac
{
\int_{\mathcal{M}_T} D\tilde\phi\, e^{-S[\phi] }\mathcal{O}[\phi]
}
{
\int_{\mathcal{M}_T} D\tilde\phi\, e^{-S[\tilde\phi] }
}    
\nn\\&=
\frac
{
\int_{\mathbb{R}^N} D\zeta \det J(\zeta)\, e^{-S[\phi(\zeta] }\mathcal{O}[\phi(\zeta)]
}
{
\int_{\mathbb{R}^N} D\zeta \det J(\zeta)\,  e^{-S[\tilde\phi] }
}
\end{align}
where  $S[\tilde\phi] $ is the analytic continuation of the action to complex values of the field, $\tilde\phi_i(\zeta_j)$ is a parameterization of the manifold $\mathcal{M}_T$ by the $N$ real variables $\zeta_j$ and $J_{ij}\equiv(\partial \tilde\phi_i/\partial\zeta_j)$ is the Jacobian relating $\tilde\phi_i$ to $\zeta_j$.

For our purposes, there are three conditions sufficient to ensure the equality of Eqs.~(\ref{eq:realint}) and (\ref{eq:comint}). First, the integrand should be holomorphic (i.e., complex analytic with no poles in the complexified domain), which is the case for most observables in quantum field theories. Second, the original and final manifolds must be homotopic --- that is, there should be a continuous family of manifolds connecting them. Lastly, the domain of integration should be compact\footnote{This condition can be relaxed for non-periodic variables if the deformation is such that 
the asymptotic behavior of the integrand guarantees the existence of the integral at all intermediate steps of the deformation.}. In the particular case of the Thirring model, all degrees of freedom are periodic, so that the original domain of integration is not $\mathbb R^N$ but $(S^1)^N = T^N$ (an N-torus) which, upon complexification, becomes $(S^1 \times \mathbb R)^N = T^N \times \mathbb R^N$.

The calculation of the integral over $\mathcal{M}_T$ requires a parameterization of the manifold. Previously the pre-image  $\zeta$ of the point $\tilde\phi$ on $\mathcal{M}_T$ was used~\cite{Alexandru:2016ejd}. In this case, the corresponding Jacobian can be calculated by solving
\be\label{eq:flow_J}
\frac{dJ_{ij}}{dt} = \overline {\frac{\partial^2 S}{\partial\zeta_i \partial\zeta_k}} \bar J_{kj},\, J(0)=\openone.
\ee Evolving $J$ according to Eq.~(\ref{eq:flow_J}) is the most computationally expensive part of the whole method.  Since $\det J$ and $S$ over $\mathcal{M}_T$  are complex, reweighting is done using:

\begin{align}\label{eq:reweight}
\langle\mathcal{O}\rangle  &= 
\frac
{
\int_{\mathbb{R}^N} D\zeta\ \det J(\zeta) e^{-S }\mathcal{O}
}
{
\int_{\mathbb{R}^N} D\zeta\ \det J(\zeta)  e^{-S }
}
\nn\\&=
\frac
{
\int_{\mathbb{R}^N} D\zeta  e^{-S_R+\re\log({\det J} ) } e^{-iS_I+i\im\log({\det J} ) }\mathcal{O}
}
{
\int_{\mathbb{R}^N} D\zeta   e^{-S +\re\log({\det J} )}
}
\nn\\&\phantom{xxxxx}\times
\frac{
\int_{\mathbb{R}^N} D\zeta   e^{-S_R +\re\log({\det J} )}
}
{
\int_{\mathbb{R}^N} D\zeta\  e^{-S_R +\re\log({\det J} ) }    e^{-iS_I+i\im\log({\det J} ) } }
\nn\\&=
\frac
{\langle \mathcal{O} e^{-iS_I+i\im\log({\det J} ) } \rangle_{S_{\rm eff}}}
{\langle e^{-iS_I+i\im\log({\det J} ) } \rangle_{S_{\rm eff}}},
\end{align} where $S_{\rm eff} = S_R - \re\log({\rm det} J)$. The phase $e^{-i S_I}$ is constant over each thimble so it typically fluctuates little over $\mathcal{M}_T$ for large enough flow time. Experience shows that $e^{i\im\log({\det J} ) } $ also  fluctuates little for problems of interest.

 The parameterization of a point $\tilde\phi(\zeta)$ of $\mathcal{M}_T$ by its pre-image $\zeta$ is problematic. This is because the regions in $\zeta$-space with large probability shrink with increasing flow time while the distance between their centers  stay fixed. The resulting probability distribution is strongly multimodal and difficult to sample via a Monte Carlo chain with small updates.  The difficulty arises because a single Metropolis proposal is unlikely to ``tunnel'' between such regions and therefore is ``trapped''.

\section{Feed-forward networks}\label{sec:ffn}
 
\begin{figure} 
 \includegraphics[width=0.5\linewidth]{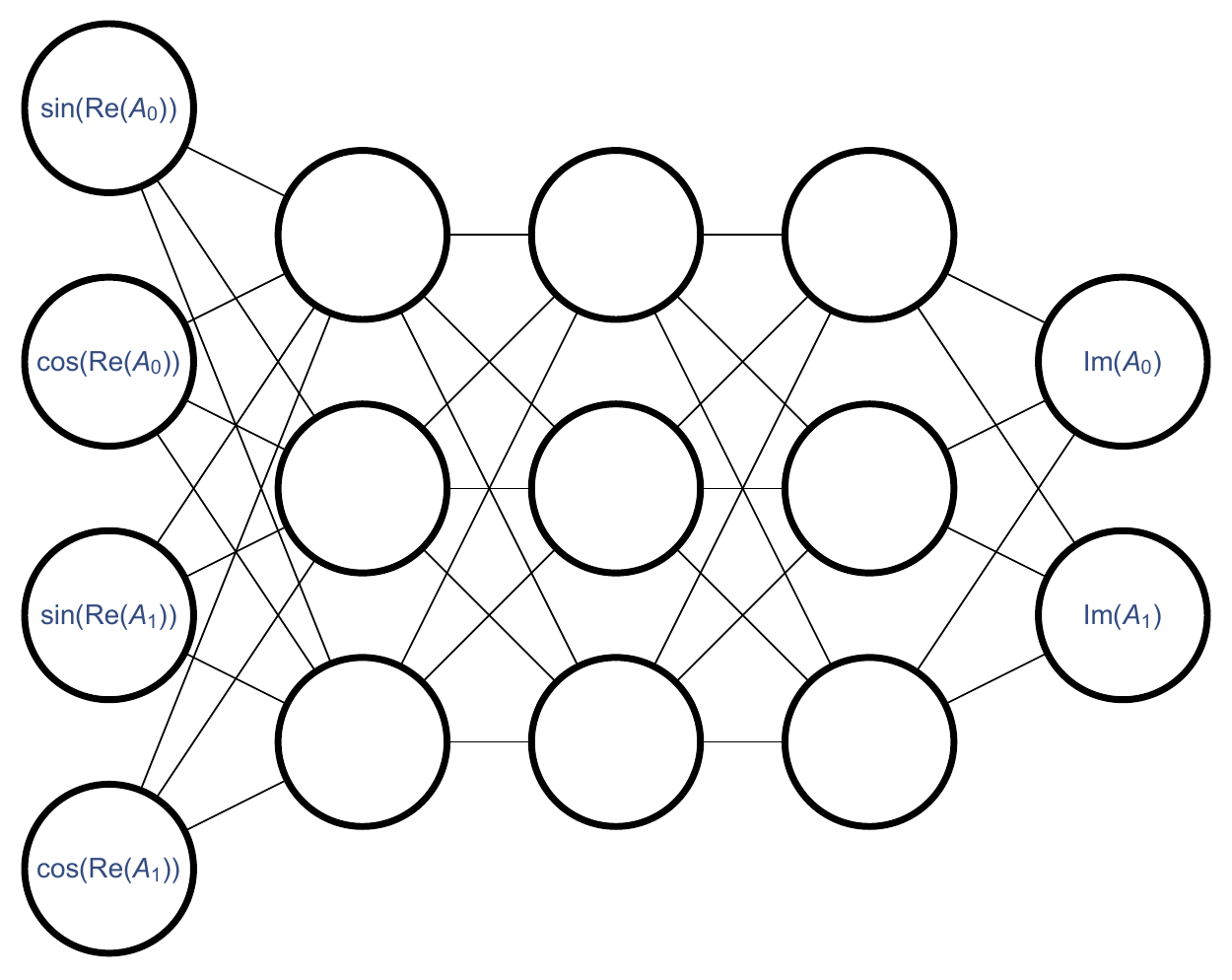}
 \caption{
 \label{fig:feed_forward}
 Graphical representation of a feed-forward network as used in the paper. The input layer is on the left, the output layer on the right.
 }
\end{figure}

In this section we summarize the use of feed-forward networks for interpolation purposes. In the next one we will detail the application of it to constructing the ``learnifold".
Feed-forward neural networks provide a family of functions particularly amenable to non-linear regression, and we use them to represent an approximation to the flowed manifold $\mathcal M_T$. A feed-forward network may be thought of as a directed graph organized into several layers which we sketch in \figu{feed_forward}. All edges point from a node in one layer to a node in the next layer. The first layer, termed the ``input layer'', has exactly as many nodes as the function has inputs. Similarly, each node of the last layer, termed the ``output layer'', corresponds to a different degree of freedom in the output of the function. There are no restrictions on the number of nodes in the intermediate (``hidden'') layers, although for simplicity, the networks we use in this paper have the same number of nodes in each hidden layer. To the edge from node $i$ to node $j$ is assigned the weight $w_{ij}$, and to each non-input node $j$ is assigned a bias $b_j$. For a fixed topology --- number of layers and nodes --- the weights and biases parameterize a family of functions. It is these weights and biases that will be adjusted when performing the non-linear regression.

Given a fixed assignment of weights and biases, the network represents a function $\vec f(\vec x)$, where $\vec x$ has as many components as the network has input nodes, and $\vec f$ has as many components as the network has output nodes. This function is evaluated in the following manner. The input values $\vec x$ are fed into the network at the input nodes. Each node $j$ in the first hidden layer computes a linear combination of these values, weighted by $w_{ij}$ and shifted by the bias $b_j$, and then applies a certain nonlinear function $\sigma$. The result becomes the output of that particular node that is then taken as the input in the nodes of the next layer. This process is repeated for each hidden layer and then the output layer, as values are ``fed forward'' through the network. Thus the value at node $j$ is given by
\be\label{eq:feed-forward}
v_j = \sigma\left(b_j + \sum_i w_{ij} v_i\right)
\ee
where the sum is taken over all nodes $i$ that have an edge leading to node $j$ (that is, all nodes in the previous layer). The values at the output nodes are taken to be the outputs of the function $\vec f(\vec x)$. The computation time required for $\vec f(\vec x)$ is linear in the number of nodes in the network.

There is considerable freedom in the choice of the nonlinear function $\sigma$. We adopt the common choice of the ``SoftPlus'' function:  
\be\label{eq:softplus}
\sigma(x) = \log\left(1+e^x\right)
\ee which asymptotically behaves like the integral of the step function, but is smoothly differentiable (making training easier). 
Because this function is bounded from below, but the function we wish to approximate may not be, we do not apply any nonlinear function at the output layer. The output values are a simple (shifted) linear combination of the values at the last hidden layer.

Our goal is to use the feed-forward network to interpolate a ``training set", that is, a set of vectors $x^{(h)}$ and their corresponding $\vec y^{(h)} = \vec f(\vec x^{(h)})$ that are assumed to be known. For that we minimize a ``cost function'' 
 \begin{align}\label{eq:cost}
C(w,b) = \frac{1}{\mathcal{N}} \sum_{h=1}^\mathcal{N} \left|\vec f_{w,b}\left(\vec x^{(h)}\right) - \vec y^{(h)}\right|.
\end{align} in relation to the biases and weights ($\mathcal{N}$ is the size of the training set).
 The minimization procedure is a simple gradient descent algorithm. For a network with $N_I$ inputs, $K$ hidden layers, $N_H$ nodes in each hidden layer, and $N_O$ outputs, there are $N_H(N_I + N_O) + (K-1)N_H^2 + KN_H + N_O$ weights and biases to be adjusted, and therefore that many dimensions to be explored during gradient descent. The gradient of the cost function is efficiently calculated through repeated use of the chain rule. Starting at the \emph{output} end of the network, we compute the gradient of the cost function with respect to the output of each node. This step is known as backpropagation.
\be\label{eq:backprop}
\frac{\partial C}{\partial v_i} = \sum_j \frac{\partial C}{\partial v_j} w_{ij} \sigma'(v_j)
\ee
Once backpropagation is complete, the gradient of the cost function with respect to the weights and biases is immediately determined:
\be\label{eq:gradient}
        \frac{\partial C}{\partial w_{ij}} = \frac{\partial C}{\partial v_j} v_i \sigma'(v_j)
        \text{\quad and \quad}
        \frac{\partial C}{\partial b_j}    = \frac{\partial C}{\partial v_j} \sigma'(v_j) \,.
\ee

Like the evaluation of the function itself, the determination of the gradient of the cost function (once the gradient with respect to the values of the output nodes is known) is linear in the number of nodes in the network.

The minimization of $C(w,b)$ is a tricky problem due to the existence of many local minima and an extensive literature is dedicated to this problem (a good review of modern gradient-descent methods is given in \cite{2016arXiv160904747R}). We use the Adaptive Moment Estimate algorithm (\textsc{Adam})~\cite{2014arXiv1412.6980K}, which was found to perform best among the methods tried. In \textsc{Adam}, the weights and biases are repeatedly updated according to the descent rule
\begin{equation}
\label{eq:adam}
	w_{s+1} = w_s - \frac{\tilde{\eta}_s}{1 - \alpha^s}(\tilde{\nabla}C)_s,
\end{equation}
where $w_s$ collectively denotes the weights and biases at iteration $s$ of the algorithm, $\tilde{\nabla}C$ is a modified gradient of the cost function with respect to $w$, and $\tilde{\eta}$ is the dynamical learning rate that determines how far along the gradient to progress. The difference in $\tilde{\nabla}C$ is the inclusion of 
``momentum'' by a decaying average of previous gradients which decreases the steps needs to reach a minima by encouraging the descent to follow the largest gradient:
\begin{equation}
  (\tilde{\nabla}C)_s=\left[(1 - \alpha) (\nabla C)_s+\alpha(\tilde{\nabla}C)_{s-1}\right].
\end{equation}
We set the weighting between the current and previous gradient to be $\alpha= 0.9$ following the suggestion of ~\cite{2016arXiv160904747R}.  The prefactor of $(1-\alpha^s)^{-1}$ in Eq.~(\ref{eq:adam}) corrects for the bias where since $w_0$ is initialized to 0, $w_s$ is biased towards remaining there.  Once $s$ is sufficiently large, this term goes to $1$ and has negligible effect.  At long times, stochastic gradient methods can oscillate around a minima, so it is useful to decrease the learning rate to reach the minima.  To decrease the learning rate, we use a dynamical learning rate is defined as
\begin{equation}
\label{eq:eta}
 \tilde{\eta}_s=\frac{\eta}{\sqrt{\frac{\tilde{v}_s}{{(1 - \beta^s)}}} + \epsilon}
\end{equation}
where $\eta = 5 \times 10^{-4}$ is a base learning rate, and $\epsilon = 10^{-8}$ is a regulator to prevent numerical instability.  Further dynamical improvement comes from using $\tilde{v}_s$ which is the variance of $\nabla C$, correcting for bias and including a momentum:
\begin{align}
	\tilde{v}_s &= (1 - \beta) \left|\nabla C\right|_s^2+\beta v_{s-1}
\end{align}
where the weighting between current and previous terms is $\beta= 0.999$ and we have again introduced into Eq.~(\ref{eq:eta}) a bias-correction factor $(1 - \beta^s)^{-1}$.
Thus, at iteration $s$ of \textsc{Adam}, we compute the gradient of the cost function $\nabla_w C$, update the estimates $(\tilde{\nabla}C)_s$ and $v_s$, and finally update the weights and biases.  We perform $10^6$ iterations to train a learnifold.

Computing the gradient of the cost function is computationally expensive due to the size of
the training set. For example $100$ configurations of $10\times 10$ lattices generate by translation a set of $10^4$ configurations (see the discussion below).  Instead,
we use a stochastic gradient descent: we approximate the gradient at each step by a sum over a small, randomly-selected batch of the configurations. We use a batch size of $25$ to begin the gradient descent, and then switch to a batch size of $200$ after $2 \times 10^5$ steps. This increase in batch size decreases the amount of stochastic noise, and the second half of the gradient descent is able to perform more fine-tuned optimizations.

\section{The learnifold}\label{sec:learnifold}

In order to avoid having to solve \eq{flow} and \eq{flow_J} at every step of the Monte Carlo chain, we will find another manifold $\mathcal{L}_T$ (the ``learnifold'') that approximates $\mathcal{M}_T$, but can be more readily computed (using a feed-forward network). Points $\tilde\phi$ on the learnifold are parameterized by points $\phi$ on the real plane:
\be
\tilde \phi_i(\phi) = \phi_i + i \tilde{f}_i(\phi)
\ee where the function $\tilde{f}$ will be constructed using the
kernel function $f$ represented by a feed-forward network (see below).

\begin{figure}
	\includegraphics[width=0.4\linewidth]{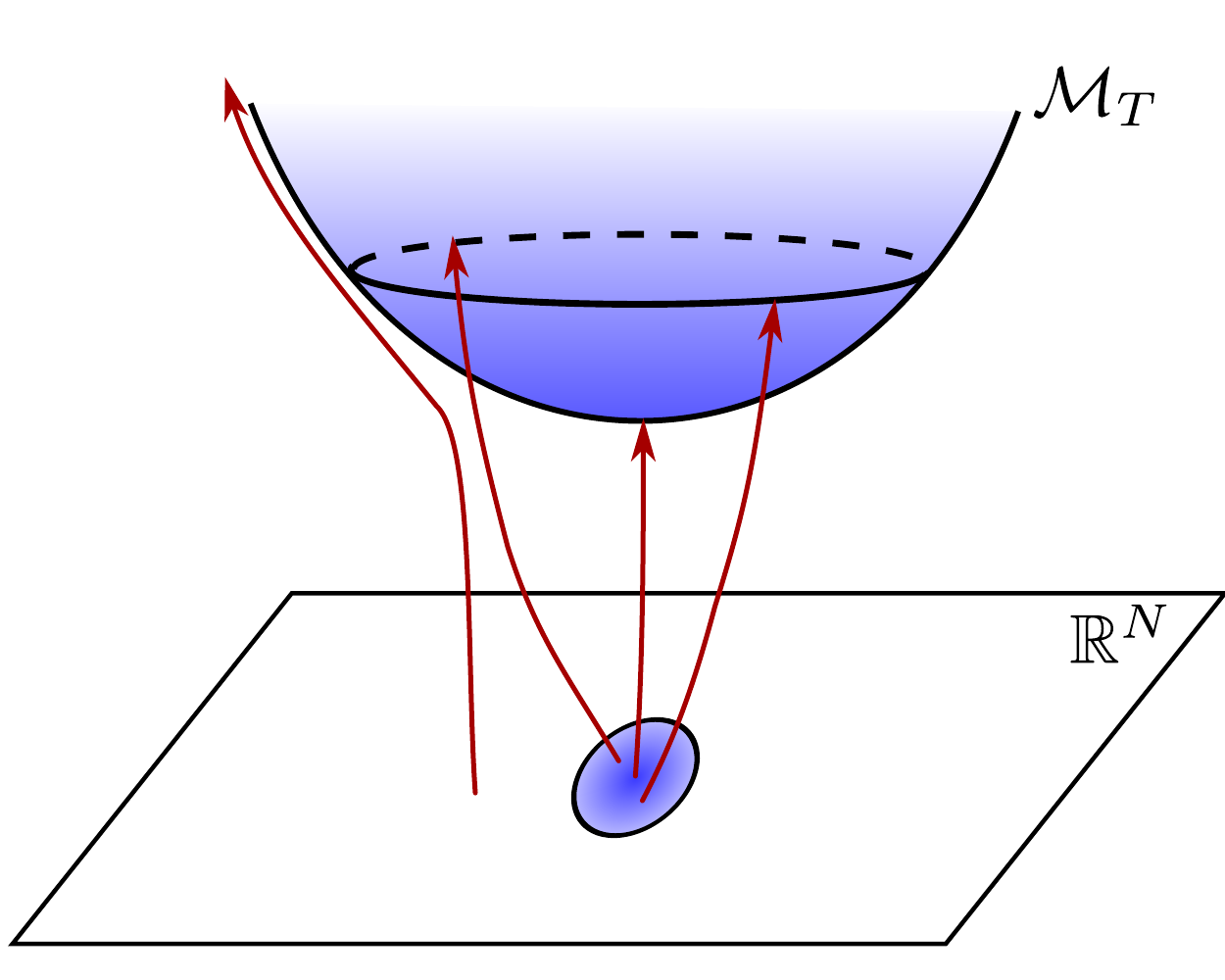}
	\includegraphics[width=0.4\linewidth]{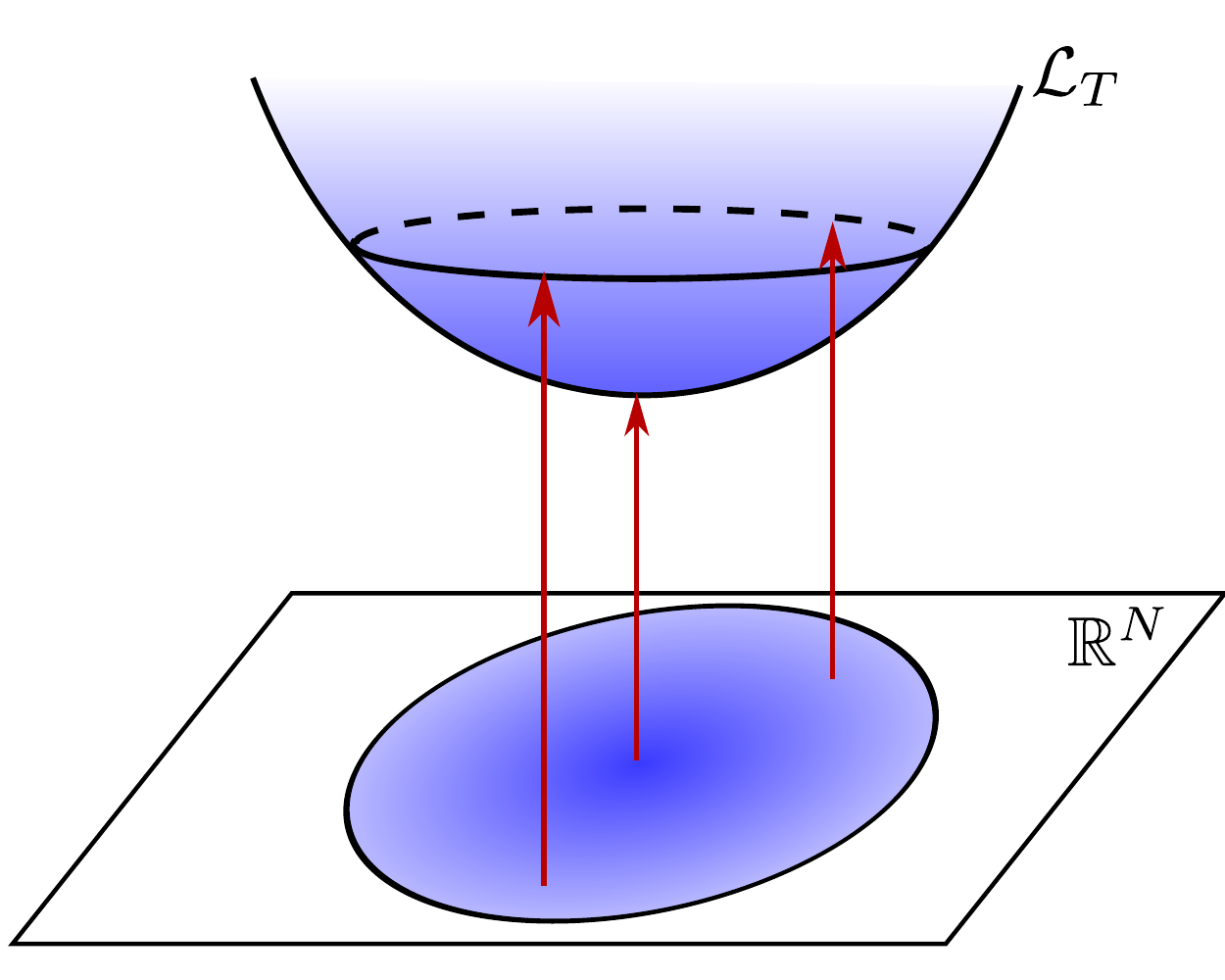}
	\caption{\label{fig:parameterization} Pictorial representation of parameterizations of manifolds. The standard generalized thimble method parameterizes (left) the manifold $\mathcal{M}_T$ by its preimage on the real plane.  This results in a large Jacobian because most regions flow into the singularities and a small region stretches. The learnifold $\mathcal{L}_T$ is parameterized (right) by its real part, so the region on the real plane is larger and the barrier between thimbles are narrower.}
\end{figure}

Besides the gain in speed from the use of a feed-forward network in place of evolution of \eq{flow}, our method differs from the one in \cite{Alexandru:2015xva,Alexandru:2016gsd,Alexandru:2015sua,Alexandru:2016ejd,Alexandru:2017lqr,Alexandru:2017oyw} by the use of a different manifold parameterization: a point in the manifold of integration is parameterized by the real part of its coordinates instead of its pre-image under the flow in \eq{flow}. This new parameterization is portrayed in \figu{parameterization}: the left-hand panel shows the parameterization arrived at from a pure-flow algorithm, and the right panel shows the parameterization from the learnifold.

This parameterization choice suffers from one drawback.  Since a point $\tilde\phi$ on the learnifold is parameterized by the real part $\phi\equiv\Re\tilde\phi$, the learnifold will necessarily have exactly one point with any given real part. In other words, the function $\tilde{f}$ defining the $\mathcal{L}_T$ is single-valued. This is a restriction on the set of manifolds that can be represented by this scheme: if the flowed manifold is such that multiple points share the same real coordinates, the class of learnifolds described here may not contain a good approximation. In practice, we find that the parts of the $\mathcal{M}_T$ that are of interest (those parts with comparatively low actions) do not behave in this way.

Despite this mild caveat, the advantages are substantial. Firstly, the parameterization of $\mathcal M_T$ by the pre-image of the flow causes a small region of parameter space (shaded blue in the figure) to map to a large region of the manifold. This results in large fluctuations of $\left|\det J\right|$, which is expensive to compute. The parameterization of $\mathcal L_T$ by the real coordinates should not lead to large stretchings. In fact, we find that $\det J \approx 1$ in practice, so that this contribution may be accounted for after-the-fact in reweighting.  The second advantage is that the parameterization reduces the multimodality problem. In $\phi$-space, regions of large probability do not shrink with the flow, and so no large gaps are created between regions that contribute significantly to the integral. A Monte Carlo chain can therefore more easily explore the relevant regions of the integration domain.

Field theoretic models of interest often have a discrete group of translational symmetries on the lattice. These symmetries are respected by the flowed manifold, and therefore should be impose on the learnifold which approximates it. 
Translation symmetry can be implemented in our setup in a simple way. 
Let $T_i$ be the lattice translation that places lattice site $i$ at the origin. 
We want that $\tilde{\phi}(T_i \phi) = T_i \tilde{\phi}(\phi)$ which requires that 
$\tilde{f}(T_i \phi) = T_i \tilde{f}(\phi)$. 
A kernel function $f:\mathbb{R}^N\to\mathbb R$ can be used to define a translational invariant
function $\tilde f:\mathbb{R}^N\to\mathbb R^N$ by:
\be
\tilde{f}_i(\phi) = f(T_i\phi)
\ee
When multiple degrees of freedom are associated with each lattice site (for the model of interest to us, there are $2$), $f$ will have that many components. 
For our case we train the kernel function $f$ to match the values of the imaginary component of the
flowed configuration at origin, that is $f(\Re\tilde\phi)\approx (\Im\tilde\phi_I)_0$, where $\tilde\phi$
are the configurations from the training set generated by flowing from $\mathbb{R}^N$.
The procedure to get $\tilde{\phi}$ on the learnifold is to start with a configuration $\phi$ in the real plane, and evaluate $f(\phi)$ to determine the imaginary part of the degrees of freedom associated to the lattice site at the origin. Then, translate the lattice so that site $i$ is moved to the origin, and evaluate $f(T_i \phi)$ again to get the imaginary part of the degrees of freedom associated to lattice site $i$. This procedure is then repeated for all possible translations of the lattice.

The inputs to the network represent the real degrees of freedom $\phi$ at each lattice site; however, in our model, these degrees of freedom are periodic. We impose this periodicity by passing to the network not $\phi$, but $\sin\phi$ and $\cos\phi$ separately --- for a model with $N$ degrees of freedom, the network will take $2N$ inputs. This is not simply an optimization: if the learnifold lacks this periodicity, it will generically belong to a different homology class from $\mathbb R^N$, that is, it will describe a manifold of integration that is not equivalent to the original domain $\mathbb R^N$.

Implementing translation invariance as described above, a point $\tilde\phi$ on the learnifold, parameterized by its real coordinates $\phi = \Re \tilde\phi$, is given by
\be\label{deformed}
\tilde\phi_i(\phi) = \phi_i + i f(T_i\phi)\text,
\ee
where $f(\phi)$, computed by a feed-forward network, outputs the imaginary parts of the degrees of freedom associated to a single lattice point.

\section{Thirring Model}\label{sec:model}

The model we use to illustrate our method, the $1+1$ dimensional Thirring model at finite chemical potential, has been studied before by the generalized thimble method~\cite{Alexandru:2016ejd}.  It is defined in the continuum by the Euclidean action
\be\label{eq:S_continuum}
S=\int d^{1+1}x\ [\bar\psi^\alpha (\slashed{\partial}+\mu \gamma_0 +m)\psi^\alpha 
+ \frac{g^2}{2N_F}\bar\psi^\alpha\gamma_\mu\psi^\alpha \bar\psi^\beta\gamma_\mu\psi^\beta],
\ee 
where the flavor indices take values $\alpha,\beta=1,\ldots,N_F$, $\mu$ is the chemical potential and the Dirac spinors $\bar\psi,\psi$ have two components. It is convenient to treat the four-fermion interaction by introducing an auxiliary vector field $A_\mu$. We use the Wilson discretization given by
\be
\!\!\!S=\sum_{x,\nu} \frac{N_F}{g^2} (1-\cos A_\nu(x))+\sum_{x,y} \bar\psi^\alpha(x) D^{W}_{xy}(A)  \psi^\alpha(y)\,,
\ee 
with
\begin{align}
D^W_{xy} = \delta_{xy} - \kappa \sum_{\nu=0,1}  
\Big[ 
 (&1-\gamma_\nu) e^{i A_\nu(x)+\mu \delta_{\nu 0}} \delta_{x+\nu, y}\nn\\
 &+ (1+\gamma_\nu) e^{-i A_\nu(x)-\mu \delta_{\nu 0}}  \delta_{x, y+\nu}
\Big], 
\end{align}
and $\kappa=1/(2m+4)$. 

The integration over the fermion fields leads to
\be
\!\!\!S=N_F 
\left(  
\frac{1}{g^2}\sum_{x,\nu} (1-\cos A_\nu(x)) - \log\det D(A)
\right)\,.
\ee 

For $\mu\neq 0$ the determinant $\det D(A)$ is not real so this model has a sign problem. In this work we use $N_F=2$.
Notice that the variables $A_\nu(x)$ are periodic so the original (real) domain of integration of the path integral is $(S^1)^N$ with $N=n L_t L_s^n $, $L_t, L_s$ being the temporal and spatial sizes of the lattice, respectively.

\section{Simulation details}\label{sec:algorithm}
Our procedure begins by using the flow-based algorithm of \cite{Alexandru:2015sua} to generate a training set, that is, a set of points on $\mathcal{M}_T$.  A quality training set need not sample the probability distribution $e^{-S_R}$ but should provide information about all of $\mathcal{M}_T$, or at least the region likely to be sampled by a long Monte-Carlo run.
A small set of training configurations sampled with $e^{-S_R}$ turns out to be insufficient. 
The learnifold generated this way develops ``pockets'', that is the distribution on it is
multimodal and the Metropolis process becomes trapped. It is then crucial to provide additional
information about $\mathcal{M}_T$.

Much freedom exists in generating this additional set: configurations normally thrown away during thermalization can be kept, and they can be generated in parallel. To cure the multimodality problem we need to include configurations on $\mathcal{M}_T$ that have large $S_R$.  For that we include in the training set configurations from an ensemble sampled from the distribution $e^{-S_{R}/\tau_i}$, with $\tau_i\geq 1$.  Sets with $\tau_i>1$ sample higher-$S_R$ regions of $\mathcal{M}_T$ than $\tau_i=1$.  We use $\tau_i=1,2$.
It should be emphasized that the ensemble generated for training is not sufficient for proper evaluation of any observable. They are not thermalized, they are highly correlated and the ones obtained with $\tau_i \neq 1$ are not distributed correctly.

Since, as discussed above, the action is translationally invariant, a single flowed configuration can be used as a total of $V$ training points, where $V$ is the space-time volume of the lattice. Thus each flowed configuration sampled is translated to $V$ other configurations. Since, in practice, the most time-consuming step of the algorithm is the generation of configurations on $\mathcal M_T$, this multiplication of the training set is critical in making this algorithm practical.

Once the training configurations are obtained, the feed-forward network is trained by minimizing the cost function. We use a network with $3$ hidden layers, each consisting of $10$ nodes. This choice is somewhat arbitrary and we have not yet fully investigated the behavior of the algorithm as the number of nodes and layers is changed.  The training is accomplished by performing stochastic gradient descent to minimize the weights $w_{ij}$ and biases $b_i$ with respect to the cost function.  Specializing Eq.~\ref{eq:cost} to our specific case:
\begin{align}
        C(w,b) = \sum_k \bigg[&\sum_\nu\left(\Im \tilde A_\nu(0,0) - \Im A_\nu(0,0)\right)^2 \bigg]^{1/2}
\end{align}
where the sum over $k$ is taken over all training points, and the sum over $\nu$ is over the Lorentz indicies. By minimizing the $C(w,b)$, we minimize the distance between the $\mathcal{L}_T$ and $\mathcal M_T$.

Once the gradient descent is complete, $f$ is used to define $\mathcal{L}_T$ through Eq.~(\ref{deformed}). This manifold is parameterized by the real plane, and so we can perform an importance sampling on this manifold in the same manner as for a flowed manifold.

\begin{figure}[t]
 \includegraphics[width=0.48\linewidth]{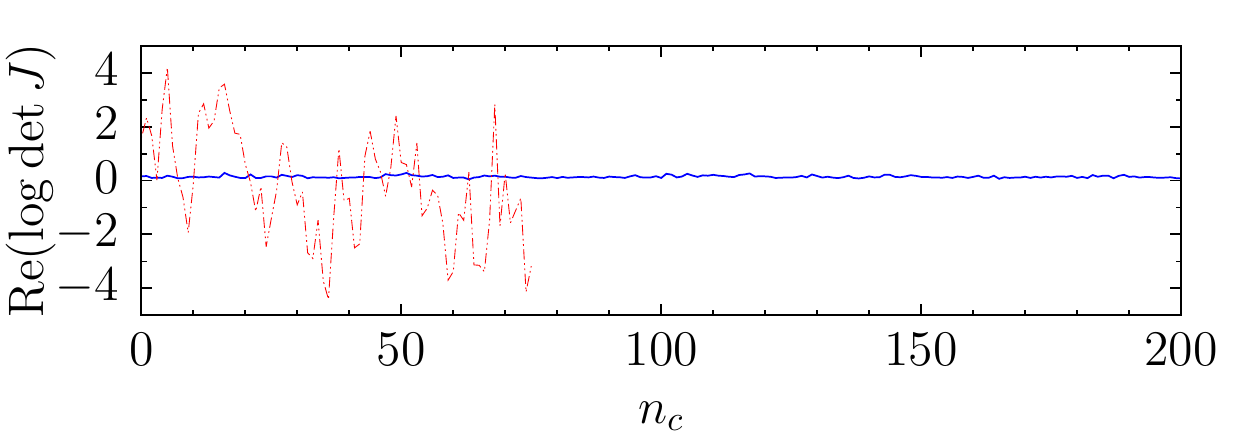}
 \includegraphics[width=0.48\linewidth]{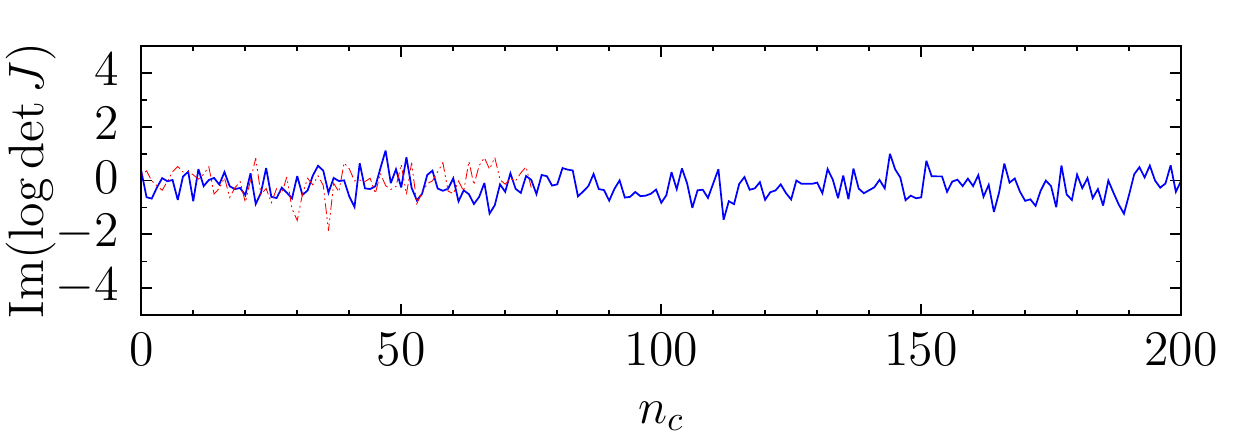}
\includegraphics[width=0.48\linewidth]{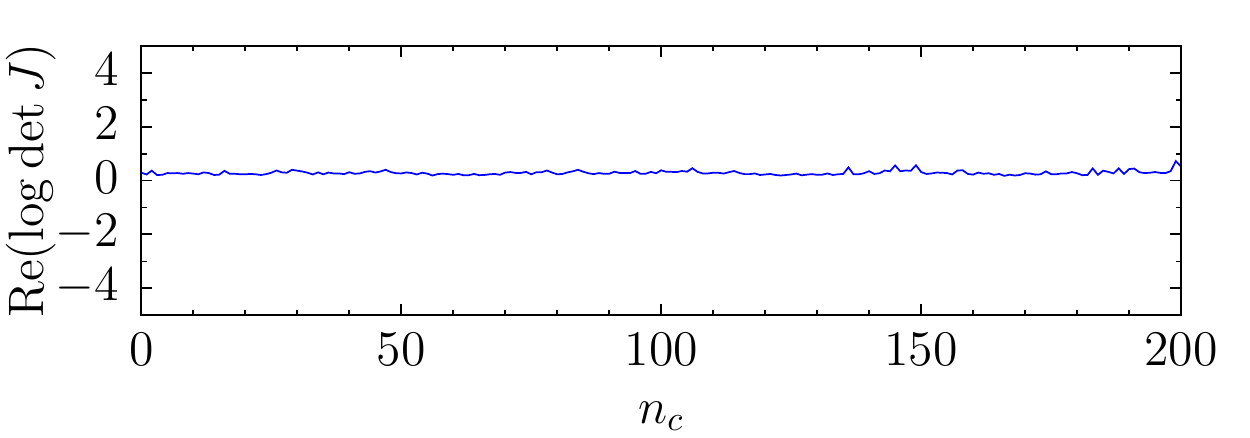}
 \includegraphics[width=0.48\linewidth]{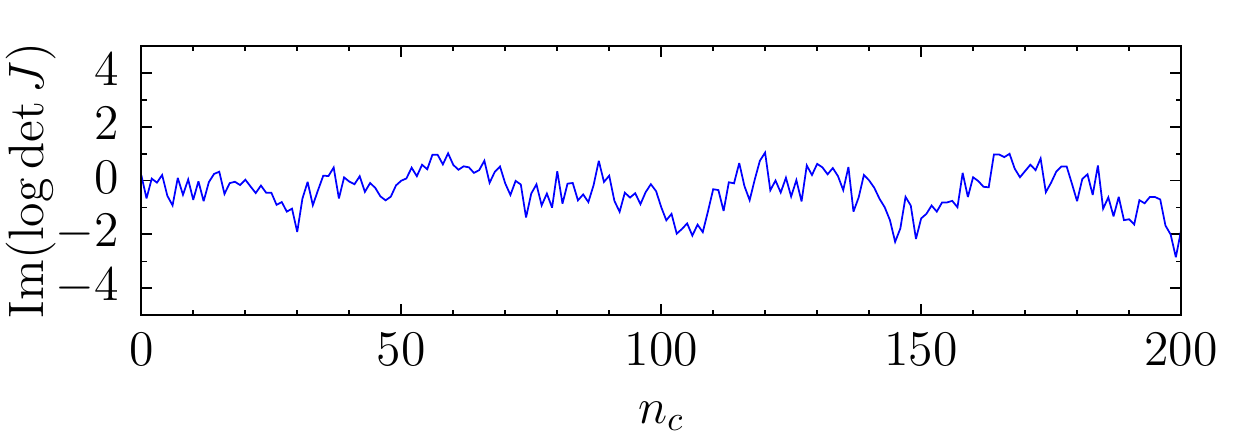}
 \caption{\label{fig:fluctuations}The Jacobian for the learnifold for the $10\times 10$
ensemble (above) and for the $40\times 10$ lattice at $\mu/m_f=2.33$ (below).
The left panels indicate the real part of the Jacobian and the right one its phase. In
the row above we compare the fluctuations of the Jacobian of the learnifold (blue) with the 
Jacobian induced by the flow (red). Note the dramatic reduction in the size of the fluctuations.
}
\end{figure}

The previous parameterization, based on deforming the domain of integration via flow, required computing the Jacobian when performing importance sampling. The $\mathcal{L}_T$ parameterization is found to result in Jacobians that have small fluctuations as can be seen from Fig.~\ref{fig:fluctuations}. Computing the Jacobian is expensive, but since its fluctuations are small it is preferable to ignore it when performing importance sampling, and include it in observables via reweighting. The Jacobian may be computed by direct application of the chain rule. In practice, it is sufficient to compute the Jacobian via finite differencing, that is, computing  $\partial f(\phi_i)/\partial\phi_j\approx (f(\phi_i+\Delta \delta_{ij})-f(\phi_i))/\Delta$ for small $\Delta$ by feeding the values $\phi_i+\Delta \delta_{ij}$ and $\phi_i$ through the network. We take this approach here.

After the network is trained and the manifold $\mathcal{L}_T$ defined by it is specified, we use the Metropolis algorithm applied to the (real) parameterizing variables $\phi_i$ and the effective action $S_{\mathrm{eff}}(\phi) = \Re S[\tilde\phi(\phi)]$, the real part of the Euclidean action.  The Jacobian and the phase of the Euclidean action are included through reweighting:
\begin{equation}
	\left<\mathcal O\right> = \frac{\left<\mathcal O e^{-iS_I + \log\det J}\right>_{S_\text{eff}}}{\left<e^{-iS_I + \log\det J}\right>_{S_\text{eff}}}
\text,
\end{equation}
where in contrast to Eq.~(\ref{eq:reweight}), the real part of the Jacobian is also included in the reweighting.
The minimum number of configurations from $\mathcal M_T$ required for training must be empirically determined and we find it to be roughly set by the number of degrees of freedom in the fit to be performed: if the network has more degrees of freedom than the number of training points available, a long training process will overfit the data, and the final product will be unusable.

\section{Results}\label{sec:results}
Although our calculations are not done particularly close to the continuum limit, we choose the bare parameters of the action so that the (renormalized) particle masses are somewhat below the lattice scale.
Two particle masses have been measured: a fermion and a boson.  The dimensionless masses of these particles, $am_f$ and $am_b$ ($a$ is the lattice spacing) respectively, are determined via fitting the large time behavior of the correlators $\langle \mathcal{O}_\alpha(t)\mathcal{O}_{\alpha}(0)^\dag\rangle$ where $\mathcal{O}_f=\psi_1$ and $\mathcal{O}_b=\bar{\psi}_i\gamma_5(\tau_3)_{ij}\psi_j$, where the subscripts indicate flavor.  In a free theory, $m_b/m_f=2$ and we use this ratio to gauge the strength of interactions, where $m_b/m_f\ll2$ implies a strongly interacting theory.  The parameters used for the simulations in this paper are $g=1.0$ and $m=-0.25$, which 
lead to $am_f=0.30(1)$ and $am_b=0.44(1)$~\cite{Alexandru:2016ejd}. We have then $m_b/m_f=1.5(2)$.  Therefore we are studying a strongly coupled theory.

The flow time used to generate training points and a range for the size of each partition of the training set is given for each lattice in \tab{training}.  
During the generation of the training set we use an estimator of the Jacobian that has been shown to track accurately the full Jacobian~\cite{Alexandru:2016lsn}. Even using the estimator
the generation of the training set is computationally expensive. For example the ensemble
used for training the $20\times 10$ learnifold at $\mu/m_f=3.83$ we use $380$ configurations with $\tau=1$
and $225$ with $\tau=2$. To generate these configurations we use no thermalization and save 
a configuration after 200 Metropolis steps. This takes about 90 CPU-hours. The training
of the network takes 24 CPU-hours. To generate the ensemble of 7220 measurements with 800 Metropolis steps between on the learnifold takes about 140 CPU-hours.  Taking the training set time as an proxy (which neglects computing the Jacobian), this set would have taken about 4100 CPU-hours using flow.  More details about the measurements can be found in Table~\ref{tab:training}.  It should be noted that for larger lattices at larger $\mu/m_f$, to achieve the same statistics more configurations are needed due to the smallness of average sign, which is reflected by a range being reported for the number of measurements.  

\begin{table}[b]
\begin{center}
\begin{tabular}
{|c| c| c c| c c c|}
\hline\hline
Lattice&$T$&\multicolumn{2}{c|}{$n_\tau$}&$n_{therm}$&$n_{cor}$& $n_{meas}$\\
&&1&2&&&\\
\hline
$10\times10$ & $0.4$ & 1000 & 1000 &15000&800 &2000\\
$20\times10$ & $0.2,0.4$ &  300-600 & 100-700&15000&800& 2000-8000\\
$40\times10$ & $0.2$ &  50-170 & 30-150&20000&800& 800-1000\\
\hline\hline
\end{tabular}
\end{center}
\caption{\label{tab:training}Training set generation and Metropolis sampling parameters.  $T$ is the flow time. Different values of $\mu/m_f$ generate flowed configurations at different rates, so for brevity we quote the size of the training set as a ranges.  $n_\tau$ is the size of the training set for each $\tau_i$, the Metropolis ``temperature''. $n_{therm}$, $n_{cor}$, and $n_{meas}$ are the thermalization, decorrelation lengths and number of measurements respectively in the Metropolis sampling.}
\end{table}

\begin{figure}[t]
 \includegraphics[width=0.48\linewidth]{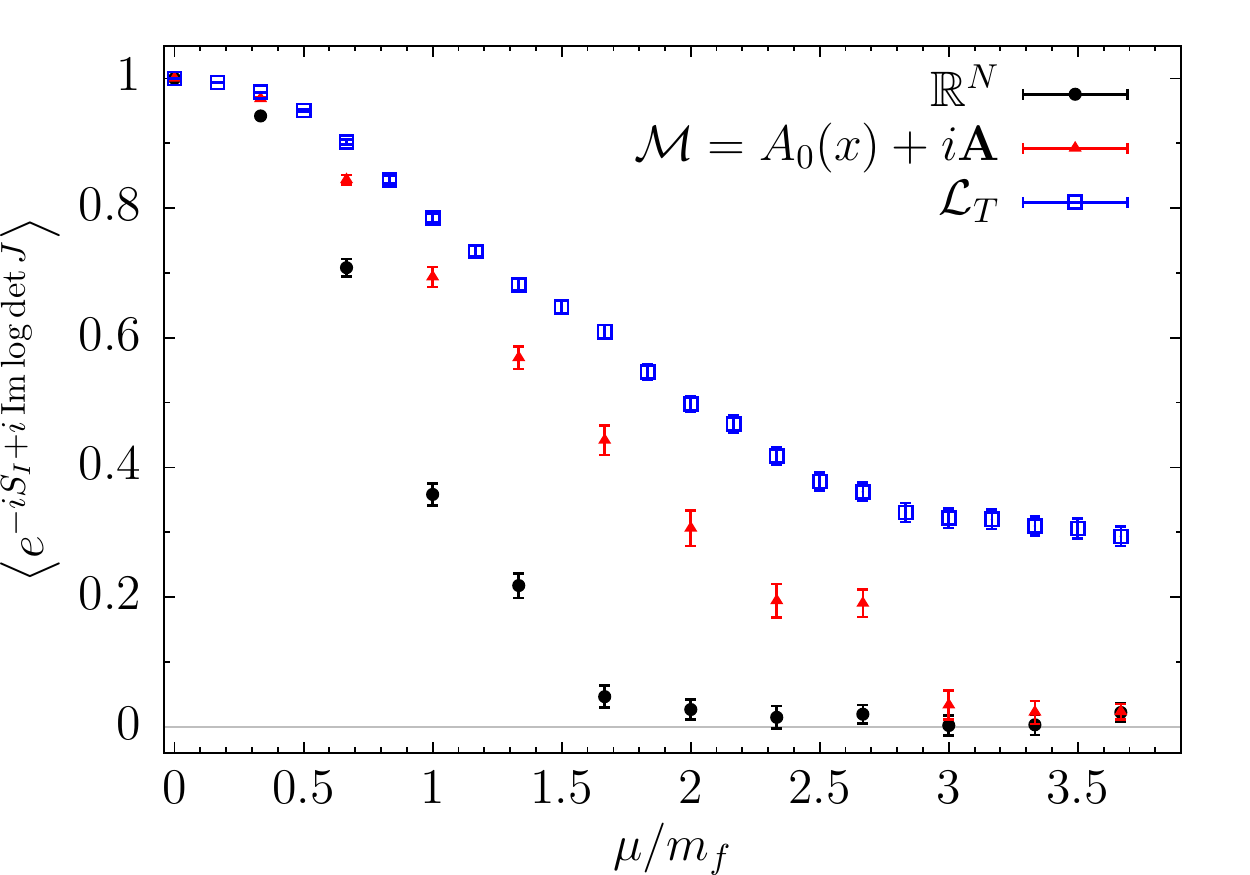}
 \includegraphics[width=0.48\linewidth]{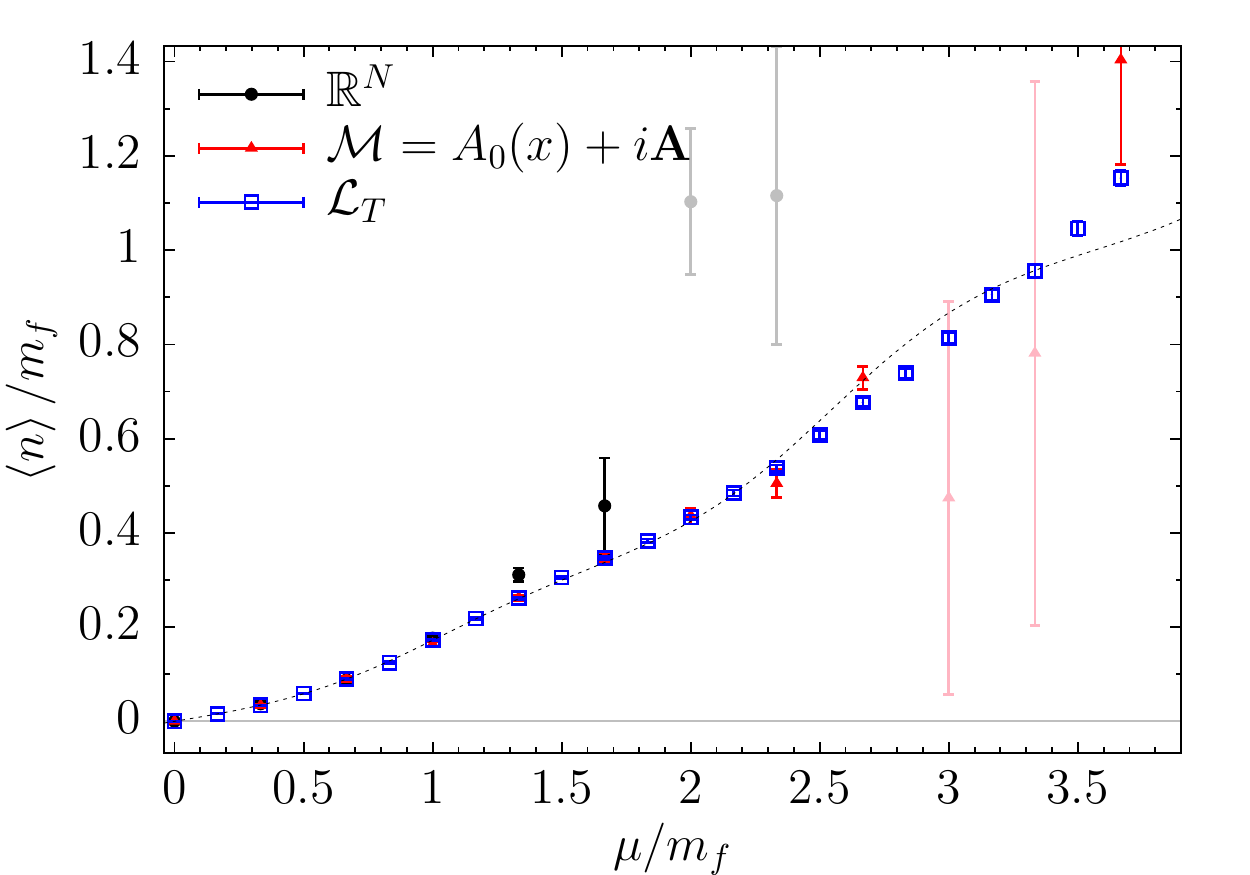}
 \includegraphics[width=0.48\linewidth]{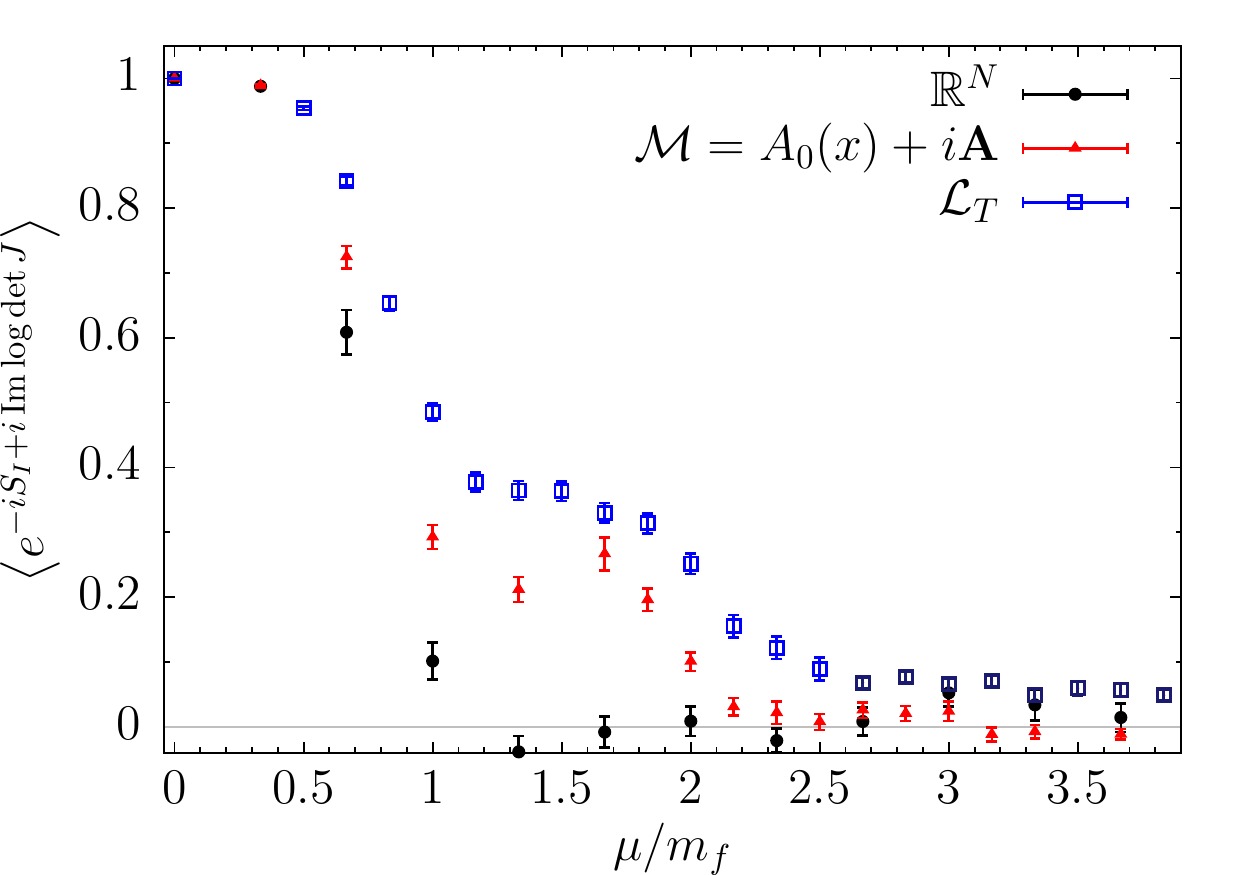}
 \includegraphics[width=0.48\linewidth]{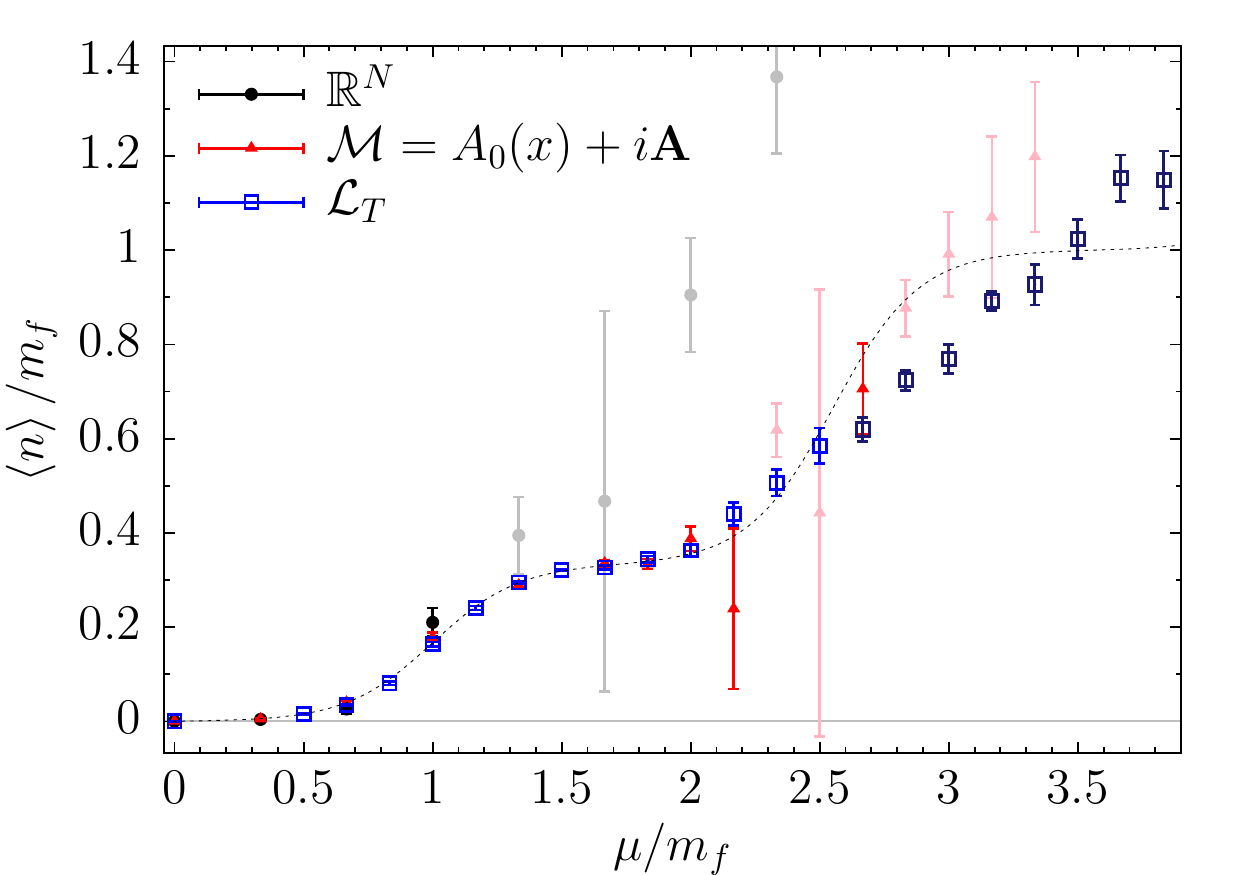}
 \includegraphics[width=0.48\linewidth]{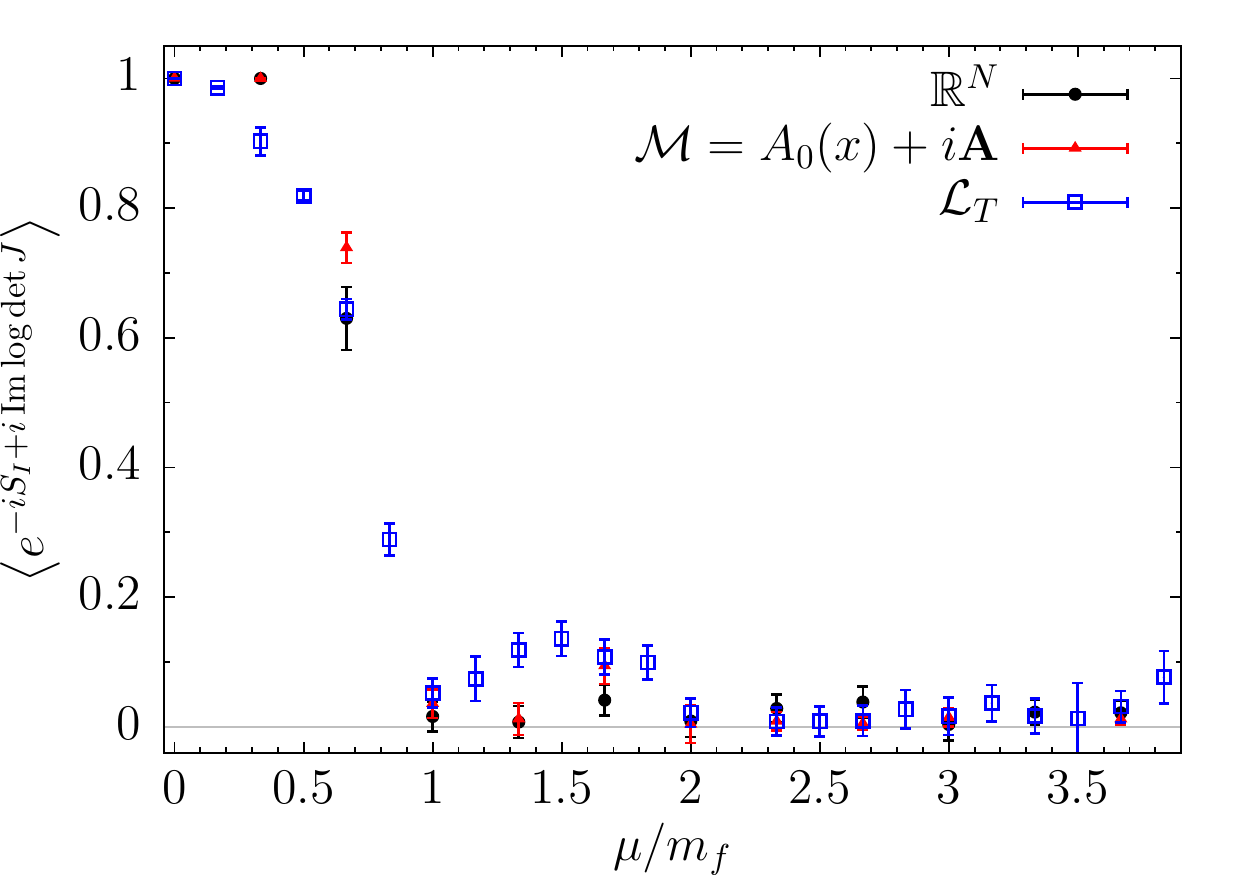}
 \includegraphics[width=0.48\linewidth]{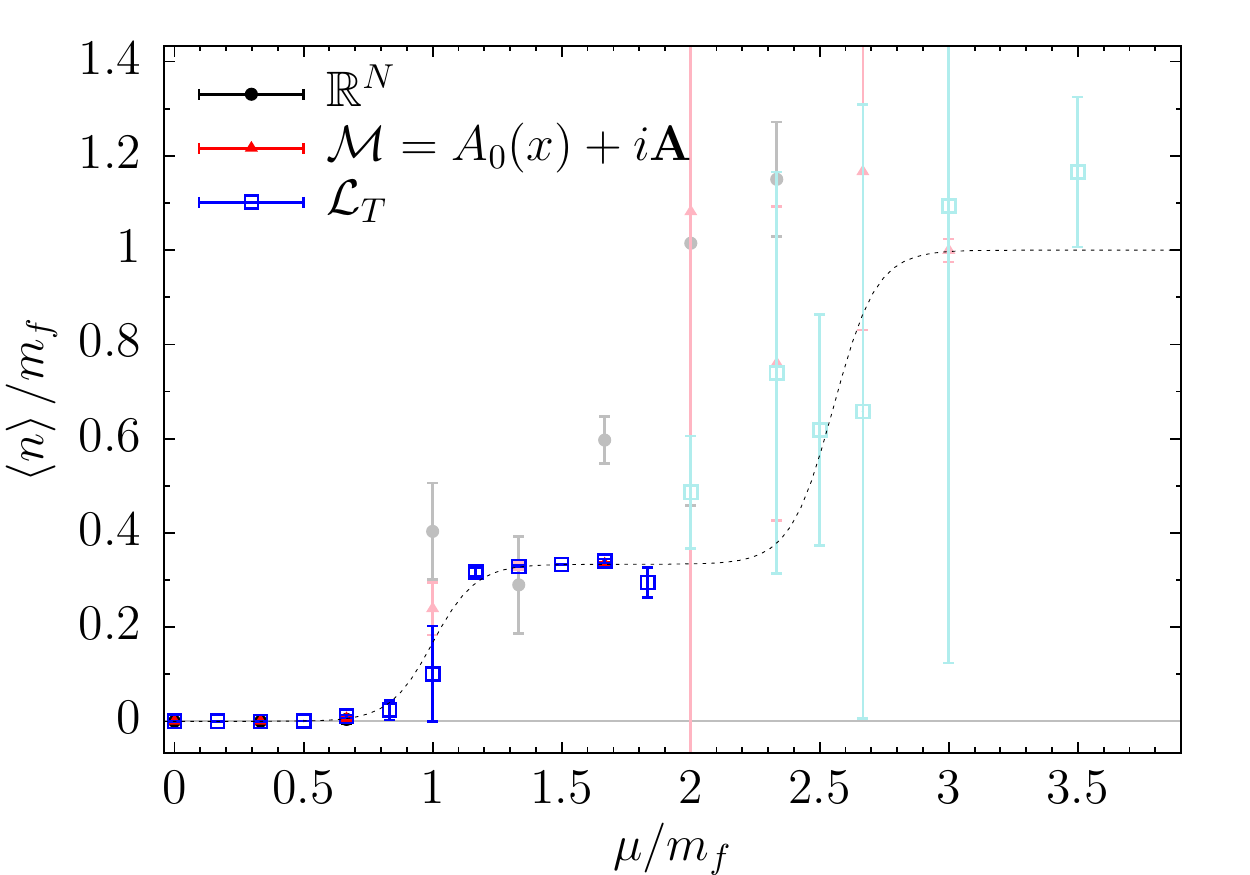}
 \caption{\label{fig:1}$\langle e^{-iS_I+i\im\log\det J}\rangle$ and $\langle n\rangle/m_f$ as a function of $\mu/m_f$ for Wilson fermions on lattices of size (top) $10\times10$, (center) $20\times10$, (bottom) $40\times10$ with $am_f=0.30(1)$. The dashed curve represents the free fermion gas with the same mass. The darker points in the $20\times 10$ graphs (middle row) correspond to a learnifold trained on ${\mathcal M}_T$ with $T_\text{flow}=0.4$
 whereas the lighter use $T_\text{flow}=0.2$.}
\end{figure}

In Fig.~\ref{fig:1}, we show for lattices of size $N_t\times N_x = 10\times 10$, $20 \times 10$, and $40\times10$ the average sign $\langle e^{-iS_I+ i\im\log\det J}\rangle$ in the left column, and in the right column the average fermion density (per flavor) $\langle n\rangle$.  The results obtained by standard reweighting techniques on $\mathbb R^N$ are shown with black circles.  As may be anticipated, the average sign drops to 0 as $\mu\approx m_f$ and reweighting becomes unfeasible.  We further plot with red triangles the results obtained by choosing as a manifold of integration the tangent plane to the main thimble (which is computationally as cheap as integration over $\mathbb R^N$).  For this manifold, the average sign drops more slowly to zero, but for sufficiently large lattices is still inadequate.  The final set of results, the blue squares, are obtained from the learnifold. We find that the average sign decrease even more slowly, extending the reach in $\mu/m_f$.  As a check, we include in these figures the analytic result for the free fermion gas (with the renormalized fermion mass) with a dashed line.  We find that for values of $\mu/m_f>2.5$, the free gas approximation becomes a poor description of the Thirring model.

As explained in \cite{Alexandru:2016ejd} the sign problem is greatly improved by simply shifting the domain of integration: $A_0(x) \rightarrow A_0(x)+i \mathbf{A}$, $A_i(x)\rightarrow A_i(x)$ for a certain real value of $\mathbf{A}$. This simple shift is enough in $1+1$ dimensions to allow for calculations in lattices of size up to $L_t\times L_s = 40\times 10$ if a staggered fermion action is used.  However, the Wilson fermion action has a worse sign problem and systems of these sizes require longer flow times which make the calculation more expensive. More importantly, the large flow times lead to the trapping of the Monte Carlo chain as discussed above.  For these reasons, Ref.~\cite{Alexandru:2016ejd} only contains simulations of the Wilson action in $10\times10$ lattices. The use of the Wilson fermion model here demonstrates the utility of machine learning in applying the generalized thimble method to larger lattices. 

On the 20$\times$10 lattice, the sign problem for $\mu/m_f\leq2.66$ is sufficiently improved for a flow time of $T=0.2$ that reliable results can be obtained.  For $\mu/m_f>2.66$, by simply flowing longer ($T=0.4$) we were able to again raise the average sign to viable levels again.  

At the colder temperatures, we demonstrate that the learnifold method is capable of reproducing the independence of observables below the threshold $\mu\approx m_f$, the so-called ``Silver Blaze'' phenomenon~\cite{Cohen:2003kd}.  This is important, because other treatments of the sign problem can fail and wash out the plateau.  In particular, we note that Lefschetz-thimble based approaches in which only the main thimble is sampled, or generalized thimble methods where trapping occurs, are likely to fail to produce the correct features and instead produce straight lines~\cite{Tanizaki:2015rda}.  We believe this is evidence that while our $\tau=1$ training sets are trapped for large flows, the learnifold parameterization reduces trapping to manageable levels in the Metropolis sampling such and the higher $\tau$ configurations give some information about the phase on additional thimbles to keep the sign problem manageable.

\section{Discussion and Prospects}\label{sec:con}
We have presented a method, based on machine learning techniques, to bypass the computationally expensive steps in the generalized thimble attack on the sign problem. The idea is that a reduced number of field configurations obtained from solving the flow equations can be used to train a feed-forward neural network that can roughly approximate the full manifold defined by the flow. The trained network defines a new manifold, the learnifold, which is equivalent to the real space for the computation of the path integral, but where the sign problem is ameliorated. The manifold defined by the network can be sampled very fast and a large number of measurements can be easily made, enough to bypass the sign problems in models where that was not previously possible. 

This paper represents a first, exploratory study of the possibility of coupling the generalized thimble method and machine learning to solve the sign problem.   Results were shown for the Wilson fermion version of the $1 + 1$ dimensional Thirring model.  As a bonus feature of the method, the parameterization we used avoids the ``trapping'' of Monte Carlo chains which plagues some calculations with large flow times.  This method is general and should be applicable to other theories of interest.  The large freedom in flow time, size and bias of the training set, and number of layers and nodes in the network together suggest that this method can be further optimized for efficiency which would extend the practical range of applicability.
 
\begin{acknowledgments}
A.A. is supported in part by the National Science Foundation CAREER grant PHY-1151648 and by U.S. Department of Energy grant DE-FG02-95ER40907. A.A. gratefully acknowledges the hospitality of the Physics Departments at the Universities of Maryland and Kentucky, and the Albert Einstein Center at the University of Bern where part of this work was carried out. 
P.F.B., H.L., and S.L.  are supported by U.S. Department of Energy under Contract No. DE-FG02-93ER-40762.
\end{acknowledgments}
\bibliographystyle{apsrev4-1}
\bibliography{thimbology}

\begin{thebibliography}{28}%
\makeatletter
\providecommand \@ifxundefined [1]{%
 \@ifx{#1\undefined}
}%
\providecommand \@ifnum [1]{%
 \ifnum #1\expandafter \@firstoftwo
 \else \expandafter \@secondoftwo
 \fi
}%
\providecommand \@ifx [1]{%
 \ifx #1\expandafter \@firstoftwo
 \else \expandafter \@secondoftwo
 \fi
}%
\providecommand \natexlab [1]{#1}%
\providecommand \enquote  [1]{``#1''}%
\providecommand \bibnamefont  [1]{#1}%
\providecommand \bibfnamefont [1]{#1}%
\providecommand \citenamefont [1]{#1}%
\providecommand \href@noop [0]{\@secondoftwo}%
\providecommand \href [0]{\begingroup \@sanitize@url \@href}%
\providecommand \@href[1]{\@@startlink{#1}\@@href}%
\providecommand \@@href[1]{\endgroup#1\@@endlink}%
\providecommand \@sanitize@url [0]{\catcode `\\12\catcode `\$12\catcode
  `\&12\catcode `\#12\catcode `\^12\catcode `\_12\catcode `\%12\relax}%
\providecommand \@@startlink[1]{}%
\providecommand \@@endlink[0]{}%
\providecommand \url  [0]{\begingroup\@sanitize@url \@url }%
\providecommand \@url [1]{\endgroup\@href {#1}{\urlprefix }}%
\providecommand \urlprefix  [0]{URL }%
\providecommand \Eprint [0]{\href }%
\providecommand \doibase [0]{http://dx.doi.org/}%
\providecommand \selectlanguage [0]{\@gobble}%
\providecommand \bibinfo  [0]{\@secondoftwo}%
\providecommand \bibfield  [0]{\@secondoftwo}%
\providecommand \translation [1]{[#1]}%
\providecommand \BibitemOpen [0]{}%
\providecommand \bibitemStop [0]{}%
\providecommand \bibitemNoStop [0]{.\EOS\space}%
\providecommand \EOS [0]{\spacefactor3000\relax}%
\providecommand \BibitemShut  [1]{\csname bibitem#1\endcsname}%
\let\auto@bib@innerbib\@empty
\bibitem [{\citenamefont {Aarts}\ and\ \citenamefont
  {Stamatescu}(2008)}]{Aarts:2008rr}%
  \BibitemOpen
  \bibfield  {author} {\bibinfo {author} {\bibfnamefont {G.}~\bibnamefont
  {Aarts}}\ and\ \bibinfo {author} {\bibfnamefont {I.-O.}\ \bibnamefont
  {Stamatescu}},\ }\href {\doibase 10.1088/1126-6708/2008/09/018} {\bibfield
  {journal} {\bibinfo  {journal} {JHEP}\ }\textbf {\bibinfo {volume} {09}},\
  \bibinfo {pages} {018} (\bibinfo {year} {2008})},\ \Eprint
  {http://arxiv.org/abs/0807.1597} {arXiv:0807.1597 [hep-lat]} \BibitemShut
  {NoStop}%
\bibitem [{\citenamefont {Langfeld}\ and\ \citenamefont
  {Lucini}(2016)}]{Langfeld:2016mct}%
  \BibitemOpen
  \bibfield  {author} {\bibinfo {author} {\bibfnamefont {K.}~\bibnamefont
  {Langfeld}}\ and\ \bibinfo {author} {\bibfnamefont {B.}~\bibnamefont
  {Lucini}},\ }\bibfield  {booktitle} {\emph {\bibinfo {booktitle}
  {{Proceedings, International Meeting Excited QCD 2016: Costa da Caparica,
  Portugal, March 6-12, 2016}}},\ }\href {\doibase 10.5506/APhysPolBSupp.9.503}
  {\bibfield  {journal} {\bibinfo  {journal} {Acta Phys. Polon. Supp.}\
  }\textbf {\bibinfo {volume} {9}},\ \bibinfo {pages} {503} (\bibinfo {year}
  {2016})},\ \Eprint {http://arxiv.org/abs/1606.03879} {arXiv:1606.03879
  [hep-lat]} \BibitemShut {NoStop}%
\bibitem [{\citenamefont {Alexandru}\ \emph {et~al.}(2005)\citenamefont
  {Alexandru}, \citenamefont {Faber}, \citenamefont {Horvath},\ and\
  \citenamefont {Liu}}]{Alexandru:2005ix}%
  \BibitemOpen
  \bibfield  {author} {\bibinfo {author} {\bibfnamefont {A.}~\bibnamefont
  {Alexandru}}, \bibinfo {author} {\bibfnamefont {M.}~\bibnamefont {Faber}},
  \bibinfo {author} {\bibfnamefont {I.}~\bibnamefont {Horvath}}, \ and\
  \bibinfo {author} {\bibfnamefont {K.-F.}\ \bibnamefont {Liu}},\ }\href
  {\doibase 10.1103/PhysRevD.72.114513} {\bibfield  {journal} {\bibinfo
  {journal} {Phys. Rev.}\ }\textbf {\bibinfo {volume} {D72}},\ \bibinfo {pages}
  {114513} (\bibinfo {year} {2005})},\ \Eprint
  {http://arxiv.org/abs/hep-lat/0507020} {arXiv:hep-lat/0507020 [hep-lat]}
  \BibitemShut {NoStop}%
\bibitem [{\citenamefont {de~Forcrand}\ and\ \citenamefont
  {Kratochvila}(2006)}]{deForcrand:2006ec}%
  \BibitemOpen
  \bibfield  {author} {\bibinfo {author} {\bibfnamefont {P.}~\bibnamefont
  {de~Forcrand}}\ and\ \bibinfo {author} {\bibfnamefont {S.}~\bibnamefont
  {Kratochvila}},\ }\bibfield  {booktitle} {\emph {\bibinfo {booktitle}
  {{Hadron physics, proceedings of the Workshop on Computational Hadron
  Physics, University of Cyprus, Nicosia, Cyprus, 14-17 September 2005}}},\
  }\href {\doibase 10.1016/j.nuclphysbps.2006.01.007} {\bibfield  {journal}
  {\bibinfo  {journal} {Nucl. Phys. Proc. Suppl.}\ }\textbf {\bibinfo {volume}
  {153}},\ \bibinfo {pages} {62} (\bibinfo {year} {2006})},\ \bibinfo {note}
  {[,62(2006)]},\ \Eprint {http://arxiv.org/abs/hep-lat/0602024}
  {arXiv:hep-lat/0602024 [hep-lat]} \BibitemShut {NoStop}%
\bibitem [{\citenamefont {Fodor}\ and\ \citenamefont
  {Katz}(2002)}]{Fodor:2001au}%
  \BibitemOpen
  \bibfield  {author} {\bibinfo {author} {\bibfnamefont {Z.}~\bibnamefont
  {Fodor}}\ and\ \bibinfo {author} {\bibfnamefont {S.~D.}\ \bibnamefont
  {Katz}},\ }\href {\doibase 10.1016/S0370-2693(02)01583-6} {\bibfield
  {journal} {\bibinfo  {journal} {Phys. Lett.}\ }\textbf {\bibinfo {volume}
  {B534}},\ \bibinfo {pages} {87} (\bibinfo {year} {2002})},\ \Eprint
  {http://arxiv.org/abs/hep-lat/0104001} {arXiv:hep-lat/0104001 [hep-lat]}
  \BibitemShut {NoStop}%
\bibitem [{\citenamefont {Allton}\ \emph {et~al.}(2002)\citenamefont {Allton},
  \citenamefont {Ejiri}, \citenamefont {Hands}, \citenamefont {Kaczmarek},
  \citenamefont {Karsch}, \citenamefont {Laermann}, \citenamefont {Schmidt},\
  and\ \citenamefont {Scorzato}}]{Allton:2002zi}%
  \BibitemOpen
  \bibfield  {author} {\bibinfo {author} {\bibfnamefont {C.~R.}\ \bibnamefont
  {Allton}}, \bibinfo {author} {\bibfnamefont {S.}~\bibnamefont {Ejiri}},
  \bibinfo {author} {\bibfnamefont {S.~J.}\ \bibnamefont {Hands}}, \bibinfo
  {author} {\bibfnamefont {O.}~\bibnamefont {Kaczmarek}}, \bibinfo {author}
  {\bibfnamefont {F.}~\bibnamefont {Karsch}}, \bibinfo {author} {\bibfnamefont
  {E.}~\bibnamefont {Laermann}}, \bibinfo {author} {\bibfnamefont
  {C.}~\bibnamefont {Schmidt}}, \ and\ \bibinfo {author} {\bibfnamefont
  {L.}~\bibnamefont {Scorzato}},\ }\href {\doibase 10.1103/PhysRevD.66.074507}
  {\bibfield  {journal} {\bibinfo  {journal} {Phys. Rev.}\ }\textbf {\bibinfo
  {volume} {D66}},\ \bibinfo {pages} {074507} (\bibinfo {year} {2002})},\
  \Eprint {http://arxiv.org/abs/hep-lat/0204010} {arXiv:hep-lat/0204010
  [hep-lat]} \BibitemShut {NoStop}%
\bibitem [{\citenamefont {Chandrasekharan}(2013)}]{Chandrasekharan:2013rpa}%
  \BibitemOpen
  \bibfield  {author} {\bibinfo {author} {\bibfnamefont {S.}~\bibnamefont
  {Chandrasekharan}},\ }\href {\doibase 10.1140/epja/i2013-13090-y} {\bibfield
  {journal} {\bibinfo  {journal} {Eur. Phys. J.}\ }\textbf {\bibinfo {volume}
  {A49}},\ \bibinfo {pages} {90} (\bibinfo {year} {2013})},\ \Eprint
  {http://arxiv.org/abs/1304.4900} {arXiv:1304.4900 [hep-lat]} \BibitemShut
  {NoStop}%
\bibitem [{\citenamefont {de~Forcrand}\ and\ \citenamefont
  {Philipsen}(2007)}]{deForcrand:2006pv}%
  \BibitemOpen
  \bibfield  {author} {\bibinfo {author} {\bibfnamefont {P.}~\bibnamefont
  {de~Forcrand}}\ and\ \bibinfo {author} {\bibfnamefont {O.}~\bibnamefont
  {Philipsen}},\ }\href {\doibase 10.1088/1126-6708/2007/01/077} {\bibfield
  {journal} {\bibinfo  {journal} {JHEP}\ }\textbf {\bibinfo {volume} {01}},\
  \bibinfo {pages} {077} (\bibinfo {year} {2007})},\ \Eprint
  {http://arxiv.org/abs/hep-lat/0607017} {arXiv:hep-lat/0607017 [hep-lat]}
  \BibitemShut {NoStop}%
\bibitem [{\citenamefont {Cristoforetti}\ \emph {et~al.}(2012)\citenamefont
  {Cristoforetti}, \citenamefont {Di~Renzo},\ and\ \citenamefont
  {Scorzato}}]{Cristoforetti:2012su}%
  \BibitemOpen
  \bibfield  {author} {\bibinfo {author} {\bibfnamefont {M.}~\bibnamefont
  {Cristoforetti}}, \bibinfo {author} {\bibfnamefont {F.}~\bibnamefont
  {Di~Renzo}}, \ and\ \bibinfo {author} {\bibfnamefont {L.}~\bibnamefont
  {Scorzato}} (\bibinfo {collaboration} {AuroraScience}),\ }\href {\doibase
  10.1103/PhysRevD.86.074506} {\bibfield  {journal} {\bibinfo  {journal} {Phys.
  Rev.}\ }\textbf {\bibinfo {volume} {D86}},\ \bibinfo {pages} {074506}
  (\bibinfo {year} {2012})},\ \Eprint {http://arxiv.org/abs/1205.3996}
  {arXiv:1205.3996 [hep-lat]} \BibitemShut {NoStop}%
\bibitem [{\citenamefont {Cristoforetti}\ \emph
  {et~al.}(2014{\natexlab{a}})\citenamefont {Cristoforetti}, \citenamefont
  {Di~Renzo}, \citenamefont {Mukherjee},\ and\ \citenamefont
  {Scorzato}}]{Cristoforetti:2013qaa}%
  \BibitemOpen
  \bibfield  {author} {\bibinfo {author} {\bibfnamefont {M.}~\bibnamefont
  {Cristoforetti}}, \bibinfo {author} {\bibfnamefont {F.}~\bibnamefont
  {Di~Renzo}}, \bibinfo {author} {\bibfnamefont {A.}~\bibnamefont {Mukherjee}},
  \ and\ \bibinfo {author} {\bibfnamefont {L.}~\bibnamefont {Scorzato}},\
  }\bibfield  {booktitle} {\emph {\bibinfo {booktitle} {{Proceedings, 31st
  International Symposium on Lattice Field Theory (Lattice 2013): Mainz,
  Germany, July 29-August 3, 2013}}},\ }\href@noop {} {\bibfield  {journal}
  {\bibinfo  {journal} {PoS}\ }\textbf {\bibinfo {volume} {LATTICE2013}},\
  \bibinfo {pages} {197} (\bibinfo {year} {2014}{\natexlab{a}})},\ \Eprint
  {http://arxiv.org/abs/1312.1052} {arXiv:1312.1052 [hep-lat]} \BibitemShut
  {NoStop}%
\bibitem [{\citenamefont {Cristoforetti}\ \emph
  {et~al.}(2014{\natexlab{b}})\citenamefont {Cristoforetti}, \citenamefont
  {Di~Renzo}, \citenamefont {Eruzzi}, \citenamefont {Mukherjee}, \citenamefont
  {Schmidt}, \citenamefont {Scorzato},\ and\ \citenamefont
  {Torrero}}]{Cristoforetti:2014gsa}%
  \BibitemOpen
  \bibfield  {author} {\bibinfo {author} {\bibfnamefont {M.}~\bibnamefont
  {Cristoforetti}}, \bibinfo {author} {\bibfnamefont {F.}~\bibnamefont
  {Di~Renzo}}, \bibinfo {author} {\bibfnamefont {G.}~\bibnamefont {Eruzzi}},
  \bibinfo {author} {\bibfnamefont {A.}~\bibnamefont {Mukherjee}}, \bibinfo
  {author} {\bibfnamefont {C.}~\bibnamefont {Schmidt}}, \bibinfo {author}
  {\bibfnamefont {L.}~\bibnamefont {Scorzato}}, \ and\ \bibinfo {author}
  {\bibfnamefont {C.}~\bibnamefont {Torrero}},\ }\href {\doibase
  10.1103/PhysRevD.89.114505} {\bibfield  {journal} {\bibinfo  {journal} {Phys.
  Rev.}\ }\textbf {\bibinfo {volume} {D89}},\ \bibinfo {pages} {114505}
  (\bibinfo {year} {2014}{\natexlab{b}})},\ \Eprint
  {http://arxiv.org/abs/1403.5637} {arXiv:1403.5637 [hep-lat]} \BibitemShut
  {NoStop}%
\bibitem [{\citenamefont {Di~Renzo}\ and\ \citenamefont
  {Eruzzi}(2015)}]{DiRenzo:2015foa}%
  \BibitemOpen
  \bibfield  {author} {\bibinfo {author} {\bibfnamefont {F.}~\bibnamefont
  {Di~Renzo}}\ and\ \bibinfo {author} {\bibfnamefont {G.}~\bibnamefont
  {Eruzzi}},\ }\href {\doibase 10.1103/PhysRevD.92.085030} {\bibfield
  {journal} {\bibinfo  {journal} {Phys. Rev.}\ }\textbf {\bibinfo {volume}
  {D92}},\ \bibinfo {pages} {085030} (\bibinfo {year} {2015})},\ \Eprint
  {http://arxiv.org/abs/1507.03858} {arXiv:1507.03858 [hep-lat]} \BibitemShut
  {NoStop}%
\bibitem [{\citenamefont {Mukherjee}\ \emph {et~al.}(2013)\citenamefont
  {Mukherjee}, \citenamefont {Cristoforetti},\ and\ \citenamefont
  {Scorzato}}]{Mukherjee:2013aga}%
  \BibitemOpen
  \bibfield  {author} {\bibinfo {author} {\bibfnamefont {A.}~\bibnamefont
  {Mukherjee}}, \bibinfo {author} {\bibfnamefont {M.}~\bibnamefont
  {Cristoforetti}}, \ and\ \bibinfo {author} {\bibfnamefont {L.}~\bibnamefont
  {Scorzato}},\ }\href {\doibase 10.1103/PhysRevD.88.051502} {\bibfield
  {journal} {\bibinfo  {journal} {Phys. Rev.}\ }\textbf {\bibinfo {volume}
  {D88}},\ \bibinfo {pages} {051502} (\bibinfo {year} {2013})},\ \Eprint
  {http://arxiv.org/abs/1308.0233} {arXiv:1308.0233 [physics.comp-ph]}
  \BibitemShut {NoStop}%
\bibitem [{\citenamefont {Fujii}\ \emph {et~al.}(2015)\citenamefont {Fujii},
  \citenamefont {Kamata},\ and\ \citenamefont {Kikukawa}}]{Fujii:2015vha}%
  \BibitemOpen
  \bibfield  {author} {\bibinfo {author} {\bibfnamefont {H.}~\bibnamefont
  {Fujii}}, \bibinfo {author} {\bibfnamefont {S.}~\bibnamefont {Kamata}}, \
  and\ \bibinfo {author} {\bibfnamefont {Y.}~\bibnamefont {Kikukawa}},\ }\href
  {\doibase 10.1007/JHEP12(2015)125, 10.1007/JHEP09(2016)172} {\bibfield
  {journal} {\bibinfo  {journal} {JHEP}\ }\textbf {\bibinfo {volume} {12}},\
  \bibinfo {pages} {125} (\bibinfo {year} {2015})},\ \bibinfo {note} {[Erratum:
  JHEP09,172(2016)]},\ \Eprint {http://arxiv.org/abs/1509.09141}
  {arXiv:1509.09141 [hep-lat]} \BibitemShut {NoStop}%
\bibitem [{\citenamefont {Fukushima}\ and\ \citenamefont
  {Tanizaki}(2015)}]{Fukushima:2015qza}%
  \BibitemOpen
  \bibfield  {author} {\bibinfo {author} {\bibfnamefont {K.}~\bibnamefont
  {Fukushima}}\ and\ \bibinfo {author} {\bibfnamefont {Y.}~\bibnamefont
  {Tanizaki}},\ }\href {\doibase 10.1093/ptep/ptv152} {\bibfield  {journal}
  {\bibinfo  {journal} {PTEP}\ }\textbf {\bibinfo {volume} {2015}},\ \bibinfo
  {pages} {111A01} (\bibinfo {year} {2015})},\ \Eprint
  {http://arxiv.org/abs/1507.07351} {arXiv:1507.07351 [hep-th]} \BibitemShut
  {NoStop}%
\bibitem [{\citenamefont {Alexandru}\ \emph
  {et~al.}(2016{\natexlab{a}})\citenamefont {Alexandru}, \citenamefont
  {Basar},\ and\ \citenamefont {Bedaque}}]{Alexandru:2015xva}%
  \BibitemOpen
  \bibfield  {author} {\bibinfo {author} {\bibfnamefont {A.}~\bibnamefont
  {Alexandru}}, \bibinfo {author} {\bibfnamefont {G.}~\bibnamefont {Basar}}, \
  and\ \bibinfo {author} {\bibfnamefont {P.}~\bibnamefont {Bedaque}},\ }\href
  {\doibase 10.1103/PhysRevD.93.014504} {\bibfield  {journal} {\bibinfo
  {journal} {Phys. Rev.}\ }\textbf {\bibinfo {volume} {D93}},\ \bibinfo {pages}
  {014504} (\bibinfo {year} {2016}{\natexlab{a}})},\ \Eprint
  {http://arxiv.org/abs/1510.03258} {arXiv:1510.03258 [hep-lat]} \BibitemShut
  {NoStop}%
\bibitem [{\citenamefont {Alexandru}\ \emph
  {et~al.}(2016{\natexlab{b}})\citenamefont {Alexandru}, \citenamefont {Basar},
  \citenamefont {Bedaque}, \citenamefont {Ridgway},\ and\ \citenamefont
  {Warrington}}]{Alexandru:2015sua}%
  \BibitemOpen
  \bibfield  {author} {\bibinfo {author} {\bibfnamefont {A.}~\bibnamefont
  {Alexandru}}, \bibinfo {author} {\bibfnamefont {G.}~\bibnamefont {Basar}},
  \bibinfo {author} {\bibfnamefont {P.~F.}\ \bibnamefont {Bedaque}}, \bibinfo
  {author} {\bibfnamefont {G.~W.}\ \bibnamefont {Ridgway}}, \ and\ \bibinfo
  {author} {\bibfnamefont {N.~C.}\ \bibnamefont {Warrington}},\ }\href
  {\doibase 10.1007/JHEP05(2016)053} {\bibfield  {journal} {\bibinfo  {journal}
  {JHEP}\ }\textbf {\bibinfo {volume} {05}},\ \bibinfo {pages} {053} (\bibinfo
  {year} {2016}{\natexlab{b}})},\ \Eprint {http://arxiv.org/abs/1512.08764}
  {arXiv:1512.08764 [hep-lat]} \BibitemShut {NoStop}%
\bibitem [{\citenamefont {Tanizaki}\ \emph {et~al.}(2016)\citenamefont
  {Tanizaki}, \citenamefont {Hidaka},\ and\ \citenamefont
  {Hayata}}]{Tanizaki:2015rda}%
  \BibitemOpen
  \bibfield  {author} {\bibinfo {author} {\bibfnamefont {Y.}~\bibnamefont
  {Tanizaki}}, \bibinfo {author} {\bibfnamefont {Y.}~\bibnamefont {Hidaka}}, \
  and\ \bibinfo {author} {\bibfnamefont {T.}~\bibnamefont {Hayata}},\ }\href
  {\doibase 10.1088/1367-2630/18/3/033002} {\bibfield  {journal} {\bibinfo
  {journal} {New J. Phys.}\ }\textbf {\bibinfo {volume} {18}},\ \bibinfo
  {pages} {033002} (\bibinfo {year} {2016})},\ \Eprint
  {http://arxiv.org/abs/1509.07146} {arXiv:1509.07146 [hep-th]} \BibitemShut
  {NoStop}%
\bibitem [{\citenamefont {Nishimura}\ and\ \citenamefont
  {Shimasaki}(2017)}]{Nishimura:2017vav}%
  \BibitemOpen
  \bibfield  {author} {\bibinfo {author} {\bibfnamefont {J.}~\bibnamefont
  {Nishimura}}\ and\ \bibinfo {author} {\bibfnamefont {S.}~\bibnamefont
  {Shimasaki}},\ }\href {\doibase 10.1007/JHEP06(2017)023} {\bibfield
  {journal} {\bibinfo  {journal} {JHEP}\ }\textbf {\bibinfo {volume} {06}},\
  \bibinfo {pages} {023} (\bibinfo {year} {2017})},\ \Eprint
  {http://arxiv.org/abs/1703.09409} {arXiv:1703.09409 [hep-lat]} \BibitemShut
  {NoStop}%
\bibitem [{\citenamefont {Alexandru}\ \emph
  {et~al.}(2017{\natexlab{a}})\citenamefont {Alexandru}, \citenamefont {Basar},
  \citenamefont {Bedaque}, \citenamefont {Ridgway},\ and\ \citenamefont
  {Warrington}}]{Alexandru:2016ejd}%
  \BibitemOpen
  \bibfield  {author} {\bibinfo {author} {\bibfnamefont {A.}~\bibnamefont
  {Alexandru}}, \bibinfo {author} {\bibfnamefont {G.}~\bibnamefont {Basar}},
  \bibinfo {author} {\bibfnamefont {P.~F.}\ \bibnamefont {Bedaque}}, \bibinfo
  {author} {\bibfnamefont {G.~W.}\ \bibnamefont {Ridgway}}, \ and\ \bibinfo
  {author} {\bibfnamefont {N.~C.}\ \bibnamefont {Warrington}},\ }\href
  {\doibase 10.1103/PhysRevD.95.014502} {\bibfield  {journal} {\bibinfo
  {journal} {Phys. Rev.}\ }\textbf {\bibinfo {volume} {D95}},\ \bibinfo {pages}
  {014502} (\bibinfo {year} {2017}{\natexlab{a}})},\ \Eprint
  {http://arxiv.org/abs/1609.01730} {arXiv:1609.01730 [hep-lat]} \BibitemShut
  {NoStop}%
\bibitem [{\citenamefont {Fukuma}\ and\ \citenamefont
  {Umeda}(2017)}]{Fukuma:2017fjq}%
  \BibitemOpen
  \bibfield  {author} {\bibinfo {author} {\bibfnamefont {M.}~\bibnamefont
  {Fukuma}}\ and\ \bibinfo {author} {\bibfnamefont {N.}~\bibnamefont {Umeda}},\
  }\href@noop {} {\  (\bibinfo {year} {2017})},\ \Eprint
  {http://arxiv.org/abs/1703.00861} {arXiv:1703.00861 [hep-lat]} \BibitemShut
  {NoStop}%
\bibitem [{\citenamefont {Alexandru}\ \emph
  {et~al.}(2017{\natexlab{b}})\citenamefont {Alexandru}, \citenamefont {Basar},
  \citenamefont {Bedaque},\ and\ \citenamefont
  {Warrington}}]{Alexandru:2017oyw}%
  \BibitemOpen
  \bibfield  {author} {\bibinfo {author} {\bibfnamefont {A.}~\bibnamefont
  {Alexandru}}, \bibinfo {author} {\bibfnamefont {G.}~\bibnamefont {Basar}},
  \bibinfo {author} {\bibfnamefont {P.~F.}\ \bibnamefont {Bedaque}}, \ and\
  \bibinfo {author} {\bibfnamefont {N.~C.}\ \bibnamefont {Warrington}},\
  }\href@noop {} {\  (\bibinfo {year} {2017}{\natexlab{b}})},\ \Eprint
  {http://arxiv.org/abs/1703.02414} {arXiv:1703.02414 [hep-lat]} \BibitemShut
  {NoStop}%
\bibitem [{\citenamefont {{Ruder}}(2016)}]{2016arXiv160904747R}%
  \BibitemOpen
  \bibfield  {author} {\bibinfo {author} {\bibfnamefont {S.}~\bibnamefont
  {{Ruder}}},\ }\href@noop {} {\bibfield  {journal} {\bibinfo  {journal} {ArXiv
  e-prints}\ } (\bibinfo {year} {2016})},\ \Eprint
  {http://arxiv.org/abs/1609.04747} {arXiv:1609.04747 [cs.LG]} \BibitemShut
  {NoStop}%
\bibitem [{\citenamefont {{Kingma}}\ and\ \citenamefont
  {{Ba}}(2014)}]{2014arXiv1412.6980K}%
  \BibitemOpen
  \bibfield  {author} {\bibinfo {author} {\bibfnamefont {D.~P.}\ \bibnamefont
  {{Kingma}}}\ and\ \bibinfo {author} {\bibfnamefont {J.}~\bibnamefont
  {{Ba}}},\ }\href@noop {} {\bibfield  {journal} {\bibinfo  {journal} {ArXiv
  e-prints}\ } (\bibinfo {year} {2014})},\ \Eprint
  {http://arxiv.org/abs/1412.6980} {arXiv:1412.6980 [cs.LG]} \BibitemShut
  {NoStop}%
\bibitem [{\citenamefont {Alexandru}\ \emph
  {et~al.}(2016{\natexlab{c}})\citenamefont {Alexandru}, \citenamefont {Basar},
  \citenamefont {Bedaque}, \citenamefont {Vartak},\ and\ \citenamefont
  {Warrington}}]{Alexandru:2016gsd}%
  \BibitemOpen
  \bibfield  {author} {\bibinfo {author} {\bibfnamefont {A.}~\bibnamefont
  {Alexandru}}, \bibinfo {author} {\bibfnamefont {G.}~\bibnamefont {Basar}},
  \bibinfo {author} {\bibfnamefont {P.~F.}\ \bibnamefont {Bedaque}}, \bibinfo
  {author} {\bibfnamefont {S.}~\bibnamefont {Vartak}}, \ and\ \bibinfo {author}
  {\bibfnamefont {N.~C.}\ \bibnamefont {Warrington}},\ }\href {\doibase
  10.1103/PhysRevLett.117.081602} {\bibfield  {journal} {\bibinfo  {journal}
  {Phys. Rev. Lett.}\ }\textbf {\bibinfo {volume} {117}},\ \bibinfo {pages}
  {081602} (\bibinfo {year} {2016}{\natexlab{c}})},\ \Eprint
  {http://arxiv.org/abs/1605.08040} {arXiv:1605.08040 [hep-lat]} \BibitemShut
  {NoStop}%
\bibitem [{\citenamefont {Alexandru}\ \emph
  {et~al.}(2017{\natexlab{c}})\citenamefont {Alexandru}, \citenamefont {Basar},
  \citenamefont {Bedaque},\ and\ \citenamefont {Ridgway}}]{Alexandru:2017lqr}%
  \BibitemOpen
  \bibfield  {author} {\bibinfo {author} {\bibfnamefont {A.}~\bibnamefont
  {Alexandru}}, \bibinfo {author} {\bibfnamefont {G.}~\bibnamefont {Basar}},
  \bibinfo {author} {\bibfnamefont {P.~F.}\ \bibnamefont {Bedaque}}, \ and\
  \bibinfo {author} {\bibfnamefont {G.~W.}\ \bibnamefont {Ridgway}},\ }\href
  {\doibase 10.1103/PhysRevD.95.114501} {\bibfield  {journal} {\bibinfo
  {journal} {Phys. Rev.}\ }\textbf {\bibinfo {volume} {D95}},\ \bibinfo {pages}
  {114501} (\bibinfo {year} {2017}{\natexlab{c}})},\ \Eprint
  {http://arxiv.org/abs/1704.06404} {arXiv:1704.06404 [hep-lat]} \BibitemShut
  {NoStop}%
\bibitem [{\citenamefont {Alexandru}\ \emph
  {et~al.}(2016{\natexlab{d}})\citenamefont {Alexandru}, \citenamefont {Basar},
  \citenamefont {Bedaque}, \citenamefont {Ridgway},\ and\ \citenamefont
  {Warrington}}]{Alexandru:2016lsn}%
  \BibitemOpen
  \bibfield  {author} {\bibinfo {author} {\bibfnamefont {A.}~\bibnamefont
  {Alexandru}}, \bibinfo {author} {\bibfnamefont {G.}~\bibnamefont {Basar}},
  \bibinfo {author} {\bibfnamefont {P.~F.}\ \bibnamefont {Bedaque}}, \bibinfo
  {author} {\bibfnamefont {G.~W.}\ \bibnamefont {Ridgway}}, \ and\ \bibinfo
  {author} {\bibfnamefont {N.~C.}\ \bibnamefont {Warrington}},\ }\href
  {\doibase 10.1103/PhysRevD.93.094514} {\bibfield  {journal} {\bibinfo
  {journal} {Phys. Rev.}\ }\textbf {\bibinfo {volume} {D93}},\ \bibinfo {pages}
  {094514} (\bibinfo {year} {2016}{\natexlab{d}})},\ \Eprint
  {http://arxiv.org/abs/1604.00956} {arXiv:1604.00956 [hep-lat]} \BibitemShut
  {NoStop}%
\bibitem [{\citenamefont {Cohen}(2003)}]{Cohen:2003kd}%
  \BibitemOpen
  \bibfield  {author} {\bibinfo {author} {\bibfnamefont {T.~D.}\ \bibnamefont
  {Cohen}},\ }\href {\doibase 10.1103/PhysRevLett.91.222001} {\bibfield
  {journal} {\bibinfo  {journal} {Phys. Rev. Lett.}\ }\textbf {\bibinfo
  {volume} {91}},\ \bibinfo {pages} {222001} (\bibinfo {year} {2003})},\
  \Eprint {http://arxiv.org/abs/hep-ph/0307089} {arXiv:hep-ph/0307089 [hep-ph]}
  \BibitemShut {NoStop}%
\end{thebibliography}%
\end{document}